\documentclass[final,3p,fleqn]{elsarticle}
\usepackage{amsmath}
\usepackage{amssymb}
\usepackage{graphicx}
\usepackage{bm}
\usepackage{color}

\newcommand{\tran}{\ensuremath{^{\mathsf T}}}
\newcommand{\kb}{\ensuremath{k_{\mathrm{B}}}}
\newcommand{\rld}[1]{{\color[rgb]{0.57,0.15,0.56} #1}}

\begin{document}

\begin{frontmatter}

\title{Discretization errors in molecular dynamics simulations with deterministic and stochastic thermostats}

\author{Ruslan L. Davidchack}
\ead{rld8@le.ac.uk}
\address{Department of Mathematics, University of Leicester, Leicester, LE1 7RH, UK}

\date{\today}
\begin{abstract}

We investigate the influence of numerical discretization errors on computed averages in a molecular dynamics simulation of TIP4P liquid water at 300\,K coupled to different deterministic (Nos\'e-Hoover and Nos\'e-Poincar\'e) and stochastic (Langevin) thermostats.  We propose a couple of simple practical approaches to estimating such errors and taking them into account when computing the averages.  We show that it is possible to obtain accurate measurements of various system quantities using step sizes of up to 70\% of the stability threshold of the integrator, which for the system of TIP4P liquid water at 300\,K corresponds to the step size of about 7\,fs.

\end{abstract}


\end{frontmatter}


\section{Introduction} \label{sec:intro}
One of the important tasks facing a practitioner of molecular dynamics (MD) starting simulation of a new system is to choose an appropriate integration time step. Since this choice is highly system dependent, even the most sophisticated and well developed MD packages leave this choice to the user.  In order to make the MD simulation efficient, the step size should be chosen as large as possible.  However, too large a step size results in the instability of the numerical integration of the equations of motion.  Strong instability leads to the ``blow up'' of the simulation, while moderate instability usually manifests itself in a drift of the measured quantities which are expected to be stationary or preserved under the exact dynamics.  Even if the numerical integrator is stable and the measured quantities appear stationary, the numerical solution of the equations of motion introduces a {\em discretization error}, so that the computed averages depend on the step size.  For relatively small step sizes, small systems, and/or short simulation runs the discretization error is masked by the statistical error of the measurement (which stems from the finite duration of the simulation run).  However, when larger time steps are used and longer simulations of larger systems are carried out, the discretization error can become much larger than the statistical one.

A particularly clear example of this error can be found in the literature on dissipative particle dynamics (DPD)~\cite{Groot97,Jakobsen05,Allen06}.  Due to the softness of the DPD interactions, the numerical integrators employed there are stable up to relatively large time steps.  However, as was recognized relatively recently, the large time steps typically employed in DPD simulations lead to the appearance of various "artifacts" in the obtained results~\cite{Jakobsen05,Allen06}: the measured kinetic temperature differs from the temperature set by the thermostat, different components of an inhomogeneous system can have different measured kinetic temperatures, kinetic and configurational temperatures are not equal, pressure profiles in spatially inhomogeneous systems are not uniform, etc.  All of these artifacts disappear when the step size is sufficiently reduced.

One of the purposes of this article is to point out that these and other observed artifacts are caused solely by the numerical discretization error, and therefore one must not be tempted to give them a physical interpretation.  For example, the observed differences in measured temperatures should not be interpreted as a violation of equipartition or as the deviation of the modeled system from equilibrium.  Similarly, the nonuniformity of the pressure profile should not be interpreted as evidence for the presence of some hidden internal forces within the system.  Instead, a suitable interpretation of these artifacts should be given within the numerical analysis framework.

The {\em backward error analysis} of numerical integrators employed in MD simulations proves very useful in this respect.  Within its framework,
the approximate numerical solution of the MD flow $\dot{z} = F(z)$ by a numerical integrator of order $r$ with step size $h$ is interpreted as the exact solution of a modified flow $\tilde{F}(z; h)$, which can be formally expressed as a series in powers of $h$:
\[ \tilde{F}(z; h) = F(z) + h^r F_{[r]}(z) + h^{r+1}F_{[r+1]}(z) + \cdots\,. \]
The terms $F_{[j]}(z)$ can be expressed in terms of $F(z)$ and its derivatives, provided $F(z)$ is sufficiently smooth~\cite{Hairer94}.  If a numerical method is time reversible, then $F_{[j]} = 0$ for odd $j$.  If the flow $F(z)$ is Hamiltonian with Hamiltonian $H(z)$ and the numerical method is symplectic, then the modified flow is also Hamiltonian with
\[ \tilde{H}(z; h) = H(z) + h^r H_{[r]}(z) + h^{r+1}H_{[r+1]}(z) + \cdots\,, \]
where the terms $H_{[j]}(z)$ can be expressed in terms of $H(z)$ and its derivatives.

Note that, even though the above series do not converge in general, it can be shown~\cite{Benettin94} that the difference between the exact flow and a modified flow described by a truncated series with a suitably chosen number of terms can be made exponentially small.  For symplectic integrators, this implies that the energy of the suitably truncated modified Hamiltonian remains almost constant for exponentially long periods of time.  A somewhat weaker conservation of the total energy in conservative systems is also exhibited by non-symplectic methods. (An extensive discussion of these issues can be found, for example, in Refs.~\cite{HairerBookIII,Reich99,Skeel09}.)

Based on these considerations it is clear that the numerical solution of a given MD flow can be interpreted as a simulation of a modified, or "shadow", system.  The reason we observe artifacts is that, while simulating the shadow system, we still measure properties of the original system.  In order for the measurements to be consistent, we would have to modify the expressions for the measured quantities to reflect the properties of the shadow system.  For example, the expression for the kinetic temperature, which is derived from the equipartition formula $\langle p_i\, \partial H/\partial p_i \rangle = \kb T$, should instead be based on the formula
$\langle p_i\,\partial \tilde{H}/ \partial p_i \rangle = \kb T$, which will lead to consistent and physically meaningful results, albeit for the shadow system.

Of course, we are not interested in the properties of the shadow system, but rather in those of the original system.  Therefore, we would like to determine how the phase space average of a quantity $A(z)$ defined for the original system is modified when we sample it with respect to the statistical distribution generated by the modified flow.  It can be shown~\cite{Bond07} that the result can also be expressed as a power series in $h$:
\begin{equation}\label{eq:Amod}
  \langle A \rangle_0 = \langle A \rangle_h + h^r\langle A_{[r]} \rangle_h +  h^{r+1}\langle A_{[r+1]} \rangle_h + \cdots\,
\end{equation}
where $\langle \cdots \rangle_0$ and $\langle \cdots \rangle_h$ denote the averages along ergodic trajectories of the original and the modified flow, respectively. Traditionally, the time step is chosen small enough to ensure that the difference $|\langle A \rangle_0 - \langle A \rangle_h|$ is smaller than the statistical error in the estimated value of $A$.  In order to be able to use larger time steps, it is necessary to estimate this difference and take it into account.

Using a combination of backward error analysis and statistical mechanics, it is possible to derive expressions for $A_{[j]}$.  These expressions can be evaluated in the MD simulations and used to correct the measurements of $A$~\cite{BondThesis}.  Alternatively~\cite{Bond07}, it is possible to derive a modified quantity $\tilde{A}$ such that $\langle\tilde{A}\rangle_h = \langle A \rangle_0 + O(h^{r'})$, $r' > r$, and thus obtain a higher order estimator for $A$ that is more accurate at larger time steps.  The difficulty in implementing such approaches is that the expressions for $A_{[j]}$ or $\tilde{A}$ can be quite complicated.  Also, they are specific to a particular molecular system and a particular numerical method, and thus have to be re-derived for each type of system and/or method.

In the study presented in this article we focus on the empirical investigation of discretization errors and their influence on the computed results in a system coupled to different deterministic (Nos\'e-Hoover and Nos\'e-Poincar\'e) and stochastic (Langevin) thermostats.  We choose to study the system of water molecules modeled as rigid bodies with the TIP4P parameters~\cite{Jorgensen83}.  This choice is motivated by the ubiquity of water modelling in the computational chemistry and biochemistry literature, as well as our interest in exploring the interplay of discretization errors for different types of degrees of freedom (i.e., translational and rotational) within the system.  We also propose a couple of practical approaches for taking into account or removing the discretization errors from the computed averages.  The first approach is based on extrapolating results from simulations with different step sizes.  The second approach is based on introducing a weighted coupling of the Nos\'e-Hoover thermostat to different types of degrees of freedom in order to remove the leading term in the discretization error for a measured quantity of interest.

The rest of the article is organized as follows.  In Section~\ref{sec:compute} we give the details of the simulated model and the expressions for the measured quantities, which include kinetic and configurational temperatures evaluated separately for translational and rotational degrees of freedom, potential energy, pressure, translational and rotational diffusion coefficients, and radial distribution functions.  In Section~\ref{sec:res} we present and discuss the results of determining discretization errors in the measured quantities for the isolated system (i.e., $NVE$ ensemble), as well as several constant temperature simulations where the system is coupled to a Nos\'e-Hoover or a Langevin thermostat.  In Section~\ref{sec:correct} we propose and investigate two possible approaches to correcting discretization errors: extrapolation and weighted thermostating.   We summarize our results and list the main observations in Section~\ref{sec:summary}.  We also provide in the Appendix a detailed description of the numerical algorithms employed in this study.

\section{Computations} \label{sec:compute}
\subsection{Simulated System}
For our investigation of the discretization errors, we have chosen to simulate the TIP4P model of water~\cite{Jorgensen83} with screened electrostatic interactions.  This is a rigid model of water molecules with four interaction sites.  The coordinates of the system of $N$ molecules consist of the center-of-mass positions $\mathbf{r} = \{r_i=(x_i, y_i, z_i)\tran \in \mathbb{R}^3, i = 1,\ldots,N \}$ and orientations expressed in the quaternion representation $\mathbf{q} = \{q_i=(q_i^0, q_i^1, q_i^2, q_i^3)\tran \in \mathbb{R}^4, i = 1,\ldots,N\}$ with $|q_i| = 1$.  (In the above definitions $\mathsf{T}$ denotes matrix transpose, so that $r_i$ and $q_i$ are defined to be column vectors.)  Following Ref.~\cite{Miller02}, we write the system Hamiltonian in the form
\begin{equation}\label{eq:Ham}
  H(\mathbf{r},\mathbf{p},\mathbf{q},\bm{\pi})=K_\mathrm{tra}(\mathbf{p})+ K_\mathrm{rot}(\mathbf{q},\bm{\pi}) + U(\mathbf{r},\mathbf{q}),
\end{equation}
with translational kinetic energy
\begin{equation}\label{eq:ktra}
  K_\mathrm{tra}(\mathbf{p}) = \sum_{i=1}^N\frac{p_i\tran p_i}{2m}
\end{equation}
and rotational kinetic energy
\begin{equation}\label{eq:krot}
  K_\mathrm{rot}(\mathbf{q},\bm{\pi}) = \tfrac{1}{8} \sum_{i=1}^{N} \pi_i\tran \mathbf{S}(q_i)\mathbf{I}^{-1}\mathbf{S}\tran(q_i)\pi_i\,,
\end{equation}
where $\mathbf{p} = \{p_i\in \mathbb{R}^{3}, i = 1,\ldots,N \}$ are the center-of-mass momenta conjugate to $\mathbf{r}$ and $\bm{\pi} = \{\pi_i \in \mathbb{R}^{4}, i = 1,\ldots,N \}$ are the angular momenta conjugate to $\mathbf{q}$.  Here
\begin{equation}\label{eq:S}
\mathbf{S}(q) = \left(\begin{array}{rrrr}
q^0 & -q^1 & -q^2 & -q^3\\
q^1 &  q^0 & -q^3 &  q^1\\
q^2 &  q^3 &  q^0 & -q^2\\
q^3 & -q^1 &  q^2 &  q^0\\
\end{array}\right)
\end{equation}
and $\mathbf{I} = \mathrm{diag}(I_0, I_{xx}, I_{yy}, I_{zz})$ is the diagonal matrix of principal moments of inertial of a molecule (the value of $I_0$ is inconsequential).

Hamilton's equations of motion are
\begin{subequations}\label{eq:HamEM}
\begin{align}
  \dot{r}_i =& \;\frac{p_i}{m} \label{eq:HamEMa}\\
  \dot{p}_i =& \;f_i         \label{eq:HamEMb}\\
  \dot{q}_i =&    \;\nabla_{\pi_i} K_\mathrm{rot}(\mathbf{q},\bm{\pi})\label{eq:HamEMc}\\
  \dot{\pi}_i =& -\!\nabla_{q_i} K_\mathrm{rot}(\mathbf{q},\bm{\pi}) + F_i\label{eq:HamEMd}
\end{align}
\end{subequations}
where $f_i = -\nabla_{r_i} U$ is the translational force acting on the centre of mass of molecule $i$ and \rld{$F_i = -\tilde{\nabla}_{q_i} U$} is the rotational force related to the torque acting on molecule $i$~\cite{Miller02}. \rld{Note that, while $\nabla _{r_i}$ and $\nabla _{q_i}$ are the usual gradients in the Cartesian coordinates in $\mathbb{R}^{3}$ and $\mathbb{R}^{4}$, respectively, $\tilde{\nabla}_{q_i}$ is the directional derivative tangent to the three-dimensional unit sphere $|q_i| = 1$.}

The potential energy $U(\mathbf{r},\mathbf{q})$ represents pairwise interaction between interaction sites within molecules:
\[ U(\mathbf{r},\mathbf{q}) = \sum_{j<i}\sum_{\alpha,\beta} u_{\alpha\beta}(|r_{i,\alpha} - r_{j,\beta}|)\,, \]
where  $r_{i,\alpha} = r_i + \mathbf{R}\tran(q_i) d_\alpha$ is the coordinate of the interaction site $\alpha$ within molecule $i$, with $d_\alpha$ being the site coordinate relative to the center of mass of a molecule in the molecule-fixed reference frame.  Here
\begin{equation}\label{eq:rotmat}
  \mathbf{R}(q) = 2\left(\begin{array}{ccc}
  (q^0)^2+(q^1)^2-\frac{1}{2} & q^1q^2 + q^0q^3 & q^1q^3 - q^0q^2\\
  q^1q^2 - q^0q^3 & (q^0)^2+(q^2)^2-\frac{1}{2} & q^2q^3 + q^0q^1\\
  q^1q^3 + q^0q^2 & q^2q^3 - q^0q^1 & (q^0)^2+(q^3)^2-\frac{1}{2}
  \end{array}\right)
\end{equation}
is the rotational matrix expressed in terms of quaternion coordinates. 
The pairwise interaction potential
\begin{equation}\label{eq:pair}
  u_{\alpha\beta}(r) = \left\{ \begin{array}{ll}
  4\epsilon[(\sigma/r)^{12}-(\sigma/r)^6]\,, & \alpha = \beta = \mathrm{O}\,,\\
  \kappa C_\alpha C_\beta\, \mathrm{erfc}(\lambda r) / r\,, & \alpha, \beta = \mathrm{H}, \mathrm{M}
\end{array}\right.
\end{equation}
is Lennard-Jones between the oxygen sites, with $\sigma = 3.15365\,$\AA\ and $\epsilon = 0.155\,$kcal/mol, and screened Coulomb between the hydrogens and the charge site M near the oxygen, where  $C_\mathrm{H} = 0.52\,e$, $C_\mathrm{M} = -1.04\,e$, $\kappa = 332\,$(kcal/mol)\AA$/e^2$, and $\lambda = 0.29\,$\AA$^{-1}$.  In order to introduce the interaction cut-off radius $r_c$, both the Lennard-Jones and the screened Coulomb interactions are multiplied by a twice continuously differentiable function~\cite{Horn04}
\[ \phi(z) = \left\{\begin{array}{ll}
  1, & z \leq 0\,,\\
  1 - 10z^3 + 15z^4 - 6z^5, & 0 < z < 1\,,\\
  0, & z \geq 1\,,
\end{array}\right. \]
where $z = (r^2 - r_m^2)/(r_c^2 - r_m^2)$.  This degree of potential smoothness is sufficient to ensure that the $h^2$ term in the modified Hamiltonian for a second order numerical integrator is well defined and thus can be used to interpret the leading order term of the discretization error.  To study higher order terms, more smoothness might be necessary~\cite{Engle05}.  We set $r_m = 9.5\,$\AA\ and $r_c = 10\,$\AA.  The screened Coulomb potential is similar to the damped potential introduced by Wolf {\em et al.}~\cite{Wolf99} and provides a better approximation to the full electrostatic interactions compared to simple truncation~\cite{Zahn02}.

The simulated system contained 1728 molecules in a well-equilibrated liquid state at 300\,K and density 989.85\,kg/m$^3$.  The simulation run for each thermostat and each step size started from the same initial state, which was further equilibrated for 20\,000 steps, and then the measurements were collected during the subsequent 200\,000 steps.

\subsection{Measured quantities}
\subsubsection{Static}
The temperature of the system is measured in several different ways: translational and rotational kinetic temperatures are given by
\[\langle \mathcal{T}_\mathrm{tk} \rangle = \frac{2\left\langle K_\mathrm{tra} \right\rangle}{3\kb (N-1)}\qquad \mbox{and} \qquad
  \langle \mathcal{T}_\mathrm{rk} \rangle = \frac{2\left\langle K_\mathrm{rot} \right\rangle}{3\kb N}\,,\]
respectively, while the total kinetic temperature is given by
\begin{equation}\label{eq:tk}
  \langle \mathcal{T}_\mathrm{k} \rangle = \frac{2\left\langle K_\mathrm{tra} + K_\mathrm{rot} \right\rangle}{\kb (6N-3)}\,.
\end{equation}
The translational and rotational {\em configurational} temperatures are measured using expressions
\[\langle \mathcal{T}_\mathrm{tc} \rangle = \frac{\left\langle\sum_{i=1}^N |\nabla_{r_i}U|^2 \right\rangle}{\kb\left\langle\sum_{i=1}^N \nabla_{r_i}^2 U\right\rangle}
  \qquad \mbox{and} \qquad
  \langle \mathcal{T}_\mathrm{rc} \rangle = \frac{\left\langle\sum_{i=1}^N |\nabla_{\Omega_i}U|^2 \right\rangle}{\kb\left\langle\sum_{i=1}^N \nabla_{\Omega_i}^2 U\right\rangle}\,, \]
respectively, where $\nabla_{\Omega_i}$ is the angular gradient operator for molecule $i$~\cite{Chialvo01}.
We also measure potential energy per molecule
\[ \langle \mathcal{U} \rangle = \frac{\langle U \rangle}{N} \]
and pressure
\[ \langle \mathcal{P} \rangle = \frac{\left\langle 2 K_\mathrm{tra} + \sum_{i=1}^N r_i\tran \nabla_{r_i}U\right\rangle}{3V}\,,\]
where $V$ is the volume of the system.  Angle brackets are used to represent the average over a simulation run.

\subsubsection{Drift}
As was discussed in the Introduction, with the increased step size, the numerical trajectory can exhibit weak instability, which manifests itself in a steady linear drift of measured quantities.  In order to estimate the drift in the quantity $A(t)$ over the time interval $[0, t_\mathrm{max}]$, we calculate the straight line least-squares fit to the data:
\[  A(t) = a_A + b_A t + \varepsilon(t) \]
and define the drift, $\delta_A$, as
\begin{equation}\label{eq:drift}
  \delta_A = \frac{b_A t_\mathrm{max}}{\sigma_A}
\end{equation}
where $\sigma_A$ is the estimated standard deviation of $\varepsilon(t)$.  If $|\delta_A| \sim 1$ then the drift is significant, and the result obtained by averaging $A(t)$ over the simulation time $t_\mathrm{max}$ cannot be considered reliable.

\begin{figure}[tb]
\begin{center}\includegraphics[width=14cm]{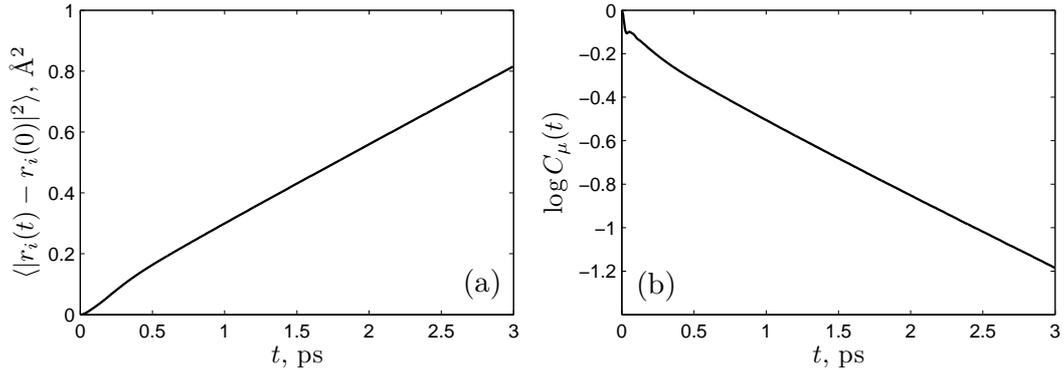}\end{center}
\caption{\label{fig:r2cmu} (a) The mean square displacement of a water molecule as a function of time measured in order to calculate the diffusion coefficient $\mathcal{D}$; (b) The dipole moment time autocorrelation function measured in order to calculate the \rld{inverse} Debye relaxation time \rld{$\tau^{-1}_\mathrm{D}$}.  The simulation is carried out using the V-NSQ integrator with 2\,fs step size.}
\end{figure}
\subsubsection{Dynamic}
Among dynamic properties of the system, we measure the diffusion coefficient $\mathcal{D}$ and the \rld{inverse} Debye relaxation time \rld{$\tau^{-1}_\mathrm{D}$}, which characterizes the rotational diffusion of water molecules.  The diffusion coefficient is computed using Einstein's relation
\begin{equation}\label{eq:eindiff}
  \mathcal{D} = \frac{1}{6}\frac{d}{dt}\langle |r_i(t)-r_i(0)|^2\rangle
\end{equation}
between the diffusion coefficient and the slope of the mean square displacement as a function of $t$.  The angle brackets denote the average over molecules and over the time origins taken at uncorrelated intervals during the simulation run.  The typical behavior of the mean square displacement as a function of time is illustrated in Figure~\ref{fig:r2cmu}(a).  For small $t$ the function is a parabola, indicating a ballistic regime, while for larger $t$ it has a linear dependence on time indicative of the diffusive regime.  The diffusion coefficient was determined from the slope of the least-squares straight line fit to this function in the range between 1 and 3\,ps.

The Debye relaxation time, $\tau_\mathrm{D}$, characterizes exponential decay of the dipole moment time autocorrelation function
\begin{equation}
  C_\mu(t) = \frac{\langle \mu_i(t) \cdot \mu_i(0) \rangle}{\langle \mu_i(0) \cdot \mu_i(0) \rangle} \sim \exp(-t/\tau_\mathrm{D})\,,
\end{equation}
where the dipole moment vector $\mu_i$, aligned with the $z$ axis in the molecule-fixed reference frame, has components proportional to the matrix elements in the bottom row of $\mathbf{R}(q_i)$ in Eq.~(\ref{eq:rotmat}).  The typical behavior of this quantity is illustrated in Figure~\ref{fig:r2cmu}(b) where $\log C_\mu(t)$ is plotted as a function of $t$.   The linear behavior of this function at larger $t$ is clearly visible\rld{, with the slope equal to $-\tau^{-1}_\mathrm{D}$}.  The \rld{inverse} Debye relaxation time was determined from the slope of the least-squares straight line fit to this function in the range between 1 and 3\,ps.

We also measure velocity and angular velocity time autocorrelation functions.  However, quantitative characterization of discretization errors in these quantities is rather problematic since, unless one uses some type of interpolation, the values are available only at integer multiples of the step size.  Therefore, we do not report the results of these measurements in this article and just note in passing that, when the diffusion coefficient $\mathcal{D}$ was calculated by integrating the measured velocity autocorrelation function using the trapezoidal rule, the results were identical within statistical errors with those obtained using Eq.~(\ref{eq:eindiff}).

\subsubsection{Structural}
To determine the influence of discretization errors on the structural properties of the system, we measure the intermolecular radial distribution functions (RDFs) for oxygen and hydrogen atoms: $g_\mathrm{OO}(r)$, $g_\mathrm{OH}(r)$, and $g_\mathrm{HH}(r)$.

\section{Results and Discussion} \label{sec:res}
In this Section we report and discuss the measurements of discretization errors in the simulation of TIP4P liquid water system coupled to various thermostats.  To solve the equations of motion (\ref{eq:HamEM}) of the isolated system (i.e., in the $NVE$ ensemble), we use a combination of velocity-Verlet integrator for translational and NO\_SQUISH integrator~\cite{Miller02} for rotational dynamics.  The combined integrator is second order, symplectic, and time reversible.   All the thermostats considered here are based on augmenting this integrator with deterministic or stochastic terms in the way that preserves the second order and, in the case of a deterministic thermostat, the time reversibility of the integrator.  Therefore, we expect that the discretization errors in the measured quantities will scale with $h$ as
\begin{equation}\label{eq:h2err}
  \langle A \rangle_h = \langle A \rangle_0 + h^2 E_A + \mathcal{O}(h^p)\,
\end{equation}
with $E_A$ measuring the size of the error for step sizes where the term $\mathcal{O}(h^p)$ can be neglected.  Here $p = 4$ for time reversible integrators and $p = 3$ otherwise.

For the RDFs, it is expected that, at every value of $r$, the leading order discretization error will also scale linearly with $h^2$, so that the results of the measurements with different step sizes can be expressed as
\begin{equation}\label{eq:h2rdf}
  \langle g_{\alpha\beta}(r) \rangle_h = \langle g_{\alpha\beta}(r) \rangle_0 + h^2 E_{g_{\alpha\beta}}(r) + \mathcal{O}(h^p)\,.
\end{equation}

\subsection{Verlet-NO\_SQUISH (V-NSQ) Integrator}
\begin{figure}[!tb]
\includegraphics[width=16cm]{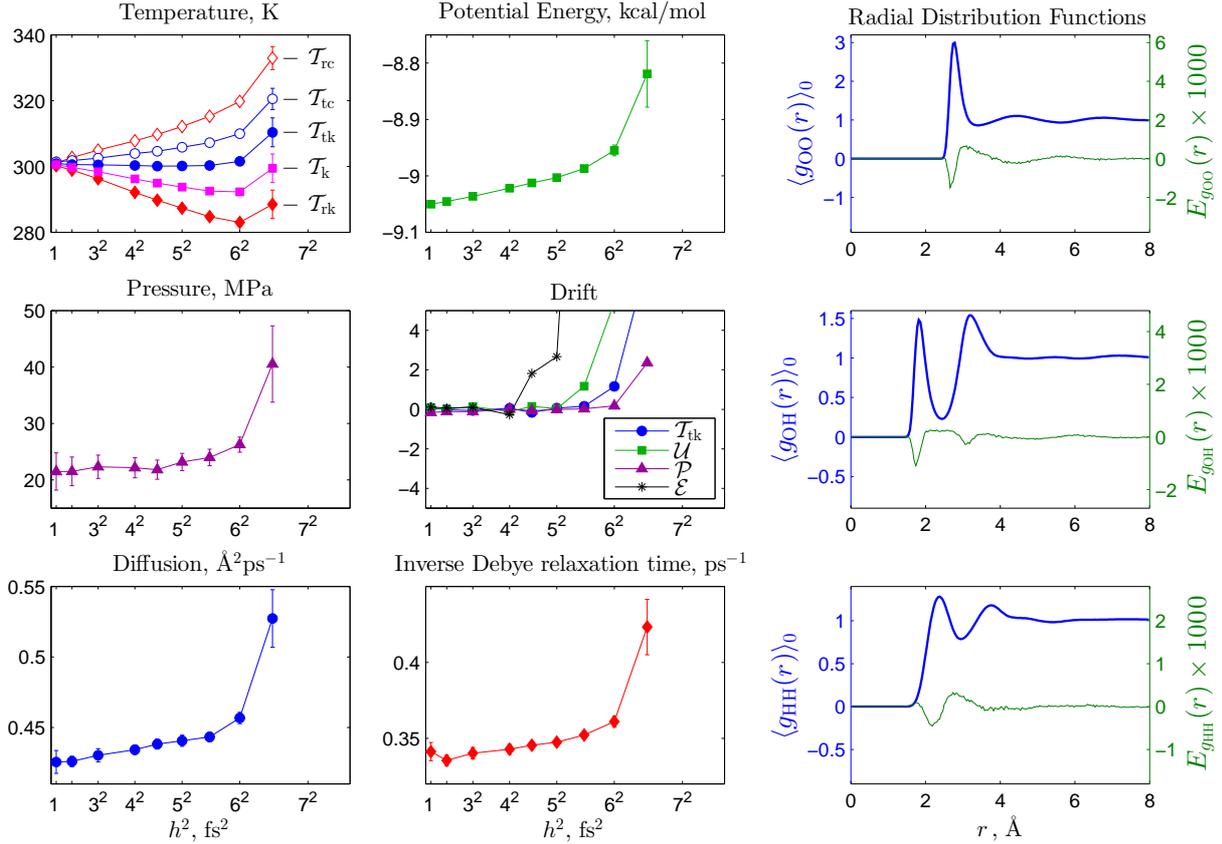}
\caption{\label{fig:nve} (Color online) Results of simulations with the V-NSQ integrator.  The left and middle columns of plots show dependence of the average measured quantities, $\langle A \rangle_h$, on the integration step size $h$, as well as the drift in the measured quantities, as defined in Eq.~(\ref{eq:drift}). The right column of plots shows results for the RDFs: $\langle g_{\alpha\beta}(r) \rangle_0$ are plotted with thick lines against the left-hand axes, while $E_{g_{\alpha\beta}}(r)$ are plotted with thin lines against the right-hand axes.  Both quantities are defined in Eq.~(\ref{eq:h2rdf}) and determined from the straight line least-squares fit to $\langle g_{\alpha\beta}(r) \rangle_h$ at each $r$ for $h \leq 5.5$\,fs.}
\end{figure}

We first report results for the constant energy ($NVE$) simulations with the Verlet-NO\_SQUISH (V-NSQ) integrator (see \ref{app:nve} for details).  The results are shown in Figure~\ref{fig:nve}.  For the static and dynamic quantities, we report $\langle A \rangle_h$ from simulations with different step sizes up to the stability threshold of the integrator.  The error bars indicate the estimated statistical errors (i.e., the 95\% confidence intervals) in the measured quantities.  As expected, the statistical errors decrease with increasing step size, which is most evident from the error bars on pressure and diffusion measurements in Figure~\ref{fig:nve}.  At the same time, the discretization errors increase.  The linear dependence of the discretization errors on $h^2$ is clearly visible and extends to the step size of about 5.5\,fs.  The integrator becomes unstable at about 7\,fs.  Prior to reaching the stability threshold, all measured quantities exhibit a drift, as defined in Eq.~(\ref{eq:drift}).  Starting with the step size of about 4.5\,fs, the total energy of the system, $\mathcal{E} = K_\mathrm{tra} + K_\mathrm{rot} + U$, is no longer conserved and starts to drift upwards, leading a gradual increase in temperature, pressure, and potential energy.  Since the total energy has much smaller dynamic fluctuations compared to other quantities, it is the most sensitive to the drift and thus it makes sense to monitor the total energy during the simulation in order to ensure the stability of the integrator.

The RDFs behave as predicted by Eq.~(\ref{eq:h2rdf}), i.e., for each value of $r$ the measured values of $\langle g_{\alpha\beta}(r) \rangle_h$ scale linearly with
$h^2$ up to about 5.5\,fs step size.  In the right column of plots in Figure~\ref{fig:nve} we show $\langle g_{\alpha\beta}(r) \rangle_0$ and $E_{g_{\alpha\beta}}(r)$, as defined in Eq.~(\ref{eq:h2rdf}) and determined from the straight line least-squares fit to $\langle g_{\alpha\beta}(r) \rangle_h$ at each $r$ for $h \leq 5.5$\,fs.  As can be seen from the plots, most of the discretization error is concentrated around the peaks of the RDFs, resulting in a reduction of the peak heights.

To facilitate a more quantitative comparison between different integrators, we list in Table~\ref{tab:nve} the values for $\langle A \rangle_0$ and $E_{A}$, as defined in Eq.~(\ref{eq:h2err}).  It shows that different measurements of the system temperature are all consistent in the limit $h \to 0$, while the discretization errors, quantified by $E_A$, are different.  In particular, the errors in rotational temperatures are larger than in translational ones, which is not surprising given the fact that the rotational motion of water molecules is faster than their translational motion.

\begin{table}[!tb]\caption{\label{tab:nve}
Results of the simulations with the V-NSQ integrator. Values for $\langle A \rangle_0$ and $E_{A}$, as defined in Eq.~(\ref{eq:h2err}), were determined from the straight line least-squares fit to $\langle A \rangle_h$ for $h \leq 5.5$\,fs.}
{\small
\begin{center}
\begin{tabular}{lcc}
\hline $A$ & $\langle A \rangle_0$ & $E_A$ \\\hline
$\mathcal{T}_\mathrm{tk}$,~K & $ 300.7(3)$ & $ -0.021(13)$\\
$\mathcal{T}_\mathrm{rk}$,~K & $ 300.9(4)$ & $ -0.540(17)$\\
$\mathcal{T}_\mathrm{k}$,~K  & $ 300.8(2)$ & $ -0.280(8)$\\
$\mathcal{T}_\mathrm{tc}$,~K & $ 300.9(5)$ & $  0.20(2)$\\
$\mathcal{T}_\mathrm{rc}$,~K & $ 300.6(5)$ & $  0.47(2)$\\
$\mathcal{U}$,~kcal/mol  & $ -9.0543(11)$ & $ 0.00212(7)$\\
$\mathcal{P}$,~MPa       & $  21.3(1.9)$ & $ 0.07(9)$\\
$\mathcal{D}$,~\AA$^2$ps$^{-1}$ & $  0.424(4)$ & $ 0.00064(16)$\\
\rld{$\tau^{-1}_\mathrm{D}$,~ps$^{-1}$}   & $  0.337(3)$ & $ 0.00045(12)$ \\\hline
\end{tabular}\end{center}}\end{table}
\begin{figure}[htb]
\includegraphics[width=16cm]{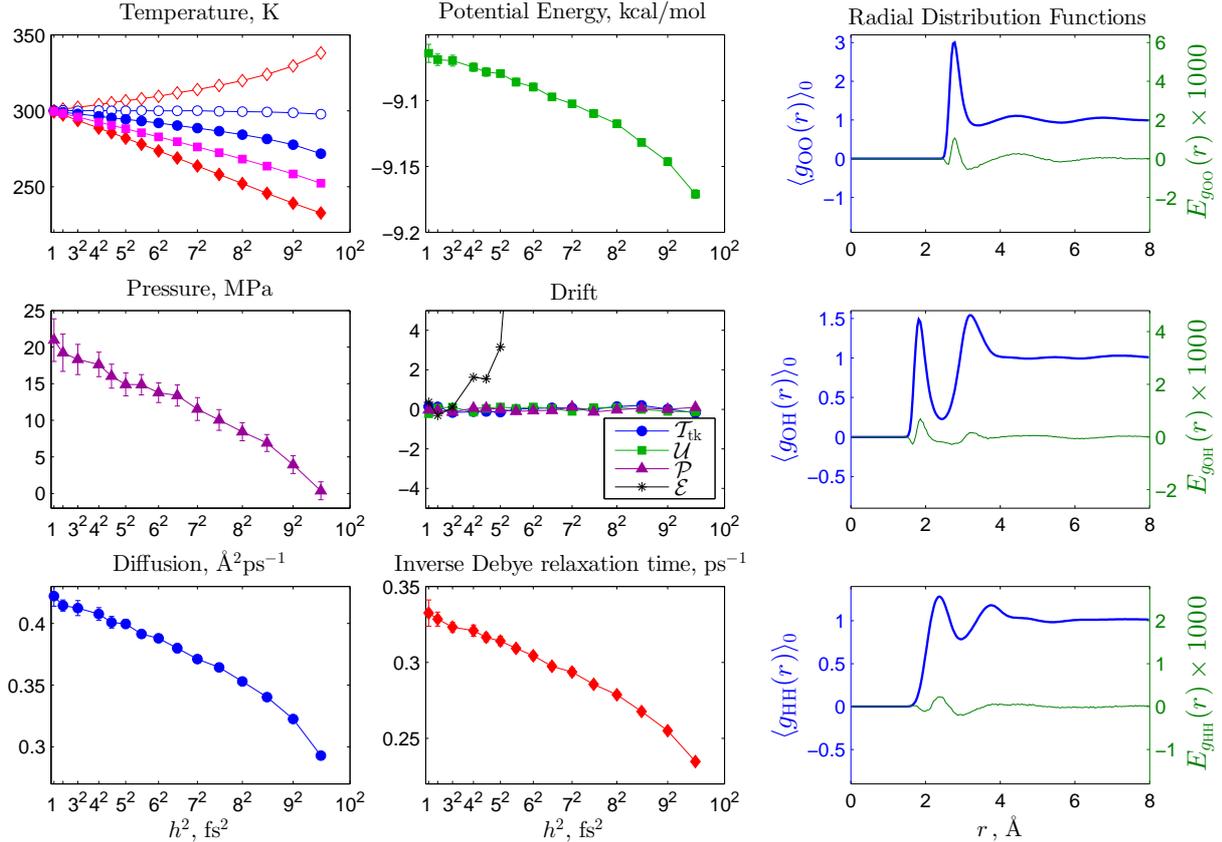}
\caption{\label{fig:nhe} (Color online) Results of simulations with the NH-E integrator.  Format of the plots is the same as in Figure~\ref{fig:nve}.  The RDF quantities $\langle g_{\alpha\beta}(r) \rangle_0$ and $E_{g_{\alpha\beta}}(r)$ were determined from the straight line least-squares fit to $\langle g_{\alpha\beta}(r) \rangle_h$ at each $r$ for $h \leq 7$\,fs.}
\end{figure}

\subsection{Nos\'e-Hoover Thermostat}
For the simulation of the system in the $NVT$ ensemble using a Nos\'e-Hoover thermostat, the equations of motion for the momenta, Eqs.~(\ref{eq:HamEMc}) and (\ref{eq:HamEMd}), are modified as follows:
\begin{align}\label{eq:NHEM}
  \dot{p}_i =& \;f_i - \xi p_i\,, \nonumber \\
  \dot{\pi}_i =& -\!\nabla_{q_i} K_\mathrm{rot}(q,\pi) + F_i - \xi \pi_i\,,
\end{align}
where $\xi$ is a scalar dynamic variable, whose evolution is described by the differential equation
\begin{equation}\label{eq:xi}
  \dot{\xi} = \frac{1}{Q}[2K_\mathrm{tra}(p) + 2K_\mathrm{rot}(q,\pi) - (6N-3)\kb T]
\end{equation}
with initial condition $\xi(0) = 0$.  Even though the modified equations are no longer Hamiltonian, they preserve the 'extended energy' of the system
\begin{equation}\label{eq:NHexen}
  \mathcal{E} = H(\mathbf{r}, \mathbf{p}, \mathbf{q}, \bm{\pi}) + \tfrac{1}{2}Q\xi^2 + (6N-3)\kb T \eta\,,
\end{equation}
where $\dot{\eta}=\xi$, $\eta(0) = 0$.  The thermostat parameter $Q$ controls the strength of thermostat coupling to the system.  Its value is usually chosen in such a way that the timescale of oscillations of $\xi(t)$ matches the natural timescale (i.e., as observed in the $NVE$ ensemble) of fluctuations of the total kinetic energy\cite{Hunenberger05}.  It is convenient to write $Q = N_\mathrm{df}\kb T \tau^2_\mathrm{NH}$, where $N_\mathrm{df}$ is the number of the thermostated degrees of freedom, and choose the value of the timescale parameter $\tau_\mathrm{NH}$.  For all Nos\'e-Hoover integrators described in this Section, as well as the Nos\'e-Poincar\'e integrator, we performed simulations with $\tau_\mathrm{NH} = 20\,$fs, 100\,fs, and 500\,fs.   However, since we have not observed any statistically significant differences in the results obtained with different values of $\tau_\mathrm{NH}$, in this Section we report results only with $\tau_\mathrm{NH} = 100\,$fs.

Eqs.~(\ref{eq:NHEM}) and (\ref{eq:xi}) are designed to control the total kinetic temperature $\mathcal{T}_\mathrm{k}$, defined in Eq.~(\ref{eq:tk}).  It is also possible to design thermostats that control only translational or rotational kinetic temperatures, or the configurational temperatures (see, for example, Ref.~\cite{Travis08}), but in this work we limit our consideration to the case of controlling $\mathcal{T}_\mathrm{k}$.

\subsubsection{Explicit Nos\'e-Hoover (NH-E) Integrator}
This is a simple second order time reversible explicit method for integrating Nos\'e-Hoover equations of motion~\cite{Holian90,Bond99} (see \ref{app:nhe} for details).  The results are shown in Figure~\ref{fig:nhe} and Table~\ref{tab:nhe}.  NH-E is more stable compared to V-NSQ, becoming unstable at step sizes of about 10\,fs.  The linear dependence of measured quantities on $h^2$ is also extended compared to V-NSQ to about 7\,fs.  The fact that NH-E controls temperature through the total kinetic energy is evident in a much smaller statistical error in $\mathcal{T}_\mathrm{k}$ compared to other temperatures.  As already mentioned, we have not observed any statistically significant differences in the results with different values of $\tau_\mathrm{NH}$, apart from the smaller estimated standard deviation in $\langle \mathcal{T}_\mathrm{k} \rangle_h$ for runs with smaller values of $\tau_\mathrm{NH}$ (i.e. stronger thermostat coupling).  Thus we obtained $\langle \mathcal{T}_\mathrm{k} \rangle_0 = 300.012(17)\,$K for $\tau_\mathrm{NH} = 20\,$fs, $300.04(9)\,$K for $\tau_\mathrm{NH} = 100\,$fs, and $300.07(19)\,$K for $\tau_\mathrm{NH} = 500\,$fs.  

Unlike in the case of the V-NSQ integrator, the drift of the extended energy $\mathcal{E}$, defined in Eq.~(\ref{eq:NHexen}), is not followed by the drift of other measured quantities.  In fact, all of the measured quantities are stationary up to the stability threshold of about 10\,fs.  Closer inspection shows that the drift in the extended energy only appears in the last term of Eq.~(\ref{eq:NHexen}) due to the drift in $\eta(t)$.  But since the value of this quantity has no influence on any of the other dynamical variables, the stationarity of the rest of the system is not perturbed.  Therefore, contrary to the common recommendations about the monitoring of the extended energy to ensure the stability of the Nos\'e-Hoover dynamics, we see that such monitoring would yield a very conservative limit on the permissible values of the time step $h$.
\begin{table}[tbp]\caption{\label{tab:nhe}
Results of the constant temperature simulations with the NH-E integrator. Values for $\langle A \rangle_0$ and $E_{A}$, as defined in Eq.~(\ref{eq:h2err}), were determined from the straight line least-squares fit to $\langle A \rangle_h$ for $h \leq 7$\,fs.}
{\small
\begin{center}
\begin{tabular}{lcc}
\hline $A$ & $\langle A \rangle_0$ & $E_A$ \\\hline
$\mathcal{T}_\mathrm{tk}$,~K & $ 299.9(2)$ & $ -0.230(7)$\\
$\mathcal{T}_\mathrm{rk}$,~K & $ 300.1(2)$ & $ -0.743(6)$\\
$\mathcal{T}_\mathrm{k}$,~K  & $ 300.04(9)$ & $ -0.486(3)$\\
$\mathcal{T}_\mathrm{tc}$,~K & $ 300.1(4)$ & $ -0.004(12)$\\
$\mathcal{T}_\mathrm{rc}$,~K & $ 299.7(3)$ & $  0.282(10)$\\
$\mathcal{U}$,~kcal/mol  & $ -9.063(3)$ & $-0.00077(9)$\\
$\mathcal{P}$,~MPa       & $  20.2(1.4)$ & $-0.18(4)$\\
$\mathcal{D}$,~\AA$^2$ps$^{-1}$ & $  0.422(3)$ & $-0.00099(8)$\\
\rld{$\tau^{-1}_\mathrm{D}$,~ps$^{-1}$}   & $  0.332(3)$ & $-0.00079(8)$\\\hline
\end{tabular}\end{center}}\end{table}

The size of the discretization errors for the RDFs is similar to that for the V-NSQ integrator, although the sign of the error is reversed, so that the peaks of $\langle g_{\alpha\beta}(r) \rangle_h$ are larger than those of $\langle g_{\alpha\beta}(r) \rangle_0$.

It is interesting to note that, even though the Nos\'e-Hoover thermostat is designed to control the total kinetic temperature of the system, this quantity also exhibits a discretization error.  The reason is that, as implemented in the NH-E integrator (see \ref{app:nhe}), the thermostat is coupled to the total kinetic energy {\em at half steps}.  Analysis of the NH-E integrator reveals the relationship between kinetic energies measured at half steps and full steps.  For simplicity, we will write NH-E for a single translational degree of freedom:
\begin{subequations}\label{eq:nhe2}
\begin{align}
  p^{n+\frac{1}{2}} =& \big(p^n + \tfrac{h}{2} f^n\big)/\big(1 + \tfrac{h}{2}\xi^n\big),\label{eq:nhea}\\
  r^{n+1} =& \;r^n + \tfrac{h}{m} p^{n+\frac{1}{2}}\,,\label{eq:nheb}\\
  \xi^{n+1} =& \;\xi^n + h\big[\big(p^{n+\frac{1}{2}}\big)^2/m - \kb T\big]/Q\,,\label{eq:nhec}\\
  p^{n+1} =& \;p^{n+\frac{1}{2}}\big(1 - \tfrac{h}{2}\xi^{n+1}\big) + \tfrac{h}{2}f^{n+1}\label{eq:nhed}
\end{align}
\end{subequations}
Assuming a well equilibrated simulation run, all measured quantities should be stationary and, in particular, $\langle \xi^{n+1} \rangle = \langle \xi^{n} \rangle$.  Taking the time average of Eq.~(\ref{eq:nhec}) we obtain
\[
  \big\langle {\big(p^{n+\frac{1}{2}}\big)^2}/{m}\big\rangle = \kb T\,,
\]
that is, the average kinetic temperature measured at half steps is equal to the thermostat temperature $T$.  To determine the difference between this kinetic temperature and the one measured at full steps, we combine
\[ p^{n+\frac{1}{2}} = \big(p^n + \tfrac{h}{2} f^n\big)/\big(1 + \tfrac{h}{2}\xi^n\big)\]
and
\[ p^{n-\frac{1}{2}} = \big(p^n - \tfrac{h}{2} f^n\big)/\big(1 - \tfrac{h}{2}\xi^n\big)\,,\]
and use again the stationarity assumption that $\big\langle {\big(p^{n+\frac{1}{2}}\big)^2} \big\rangle = \big\langle {\big(p^{n-\frac{1}{2}}\big)^2} \big\rangle$,
to obtain
\begin{equation}
  \big\langle(p^{n})^2/m\big\rangle = \kb T - \tfrac{h^2}{m}\big[\tfrac{1}{4}\big\langle(f^n)^2\big\rangle + \tfrac{3}{4}\big\langle(p^n \xi^n)^2\big\rangle - \big\langle p^n f^n \xi^n \big\rangle\big] + \mathcal{O}(h^4)
\end{equation}
This formula explains the origin of the discretization error in the measured total kinetic temperature of the system $\mathcal{T}_\mathrm{k}$.  The largest contribution to the error comes from the first term in the square brackets, which originates from the velocity splitting in the velocity-Verlet algorithm and is independent of the thermostat.  In fact, it is easy to show, using arguments similar to those above, that for the constant energy velocity-Verlet algorithm
\begin{equation}
  \big\langle(p^{n+\frac{1}{2}})^2\big\rangle = \big\langle(p^{n})^2\big\rangle + \tfrac{h^2}{4}\big\langle(f^n)^2\big\rangle\,.
\end{equation}

\begin{figure}[!tb]
\includegraphics[width=16cm]{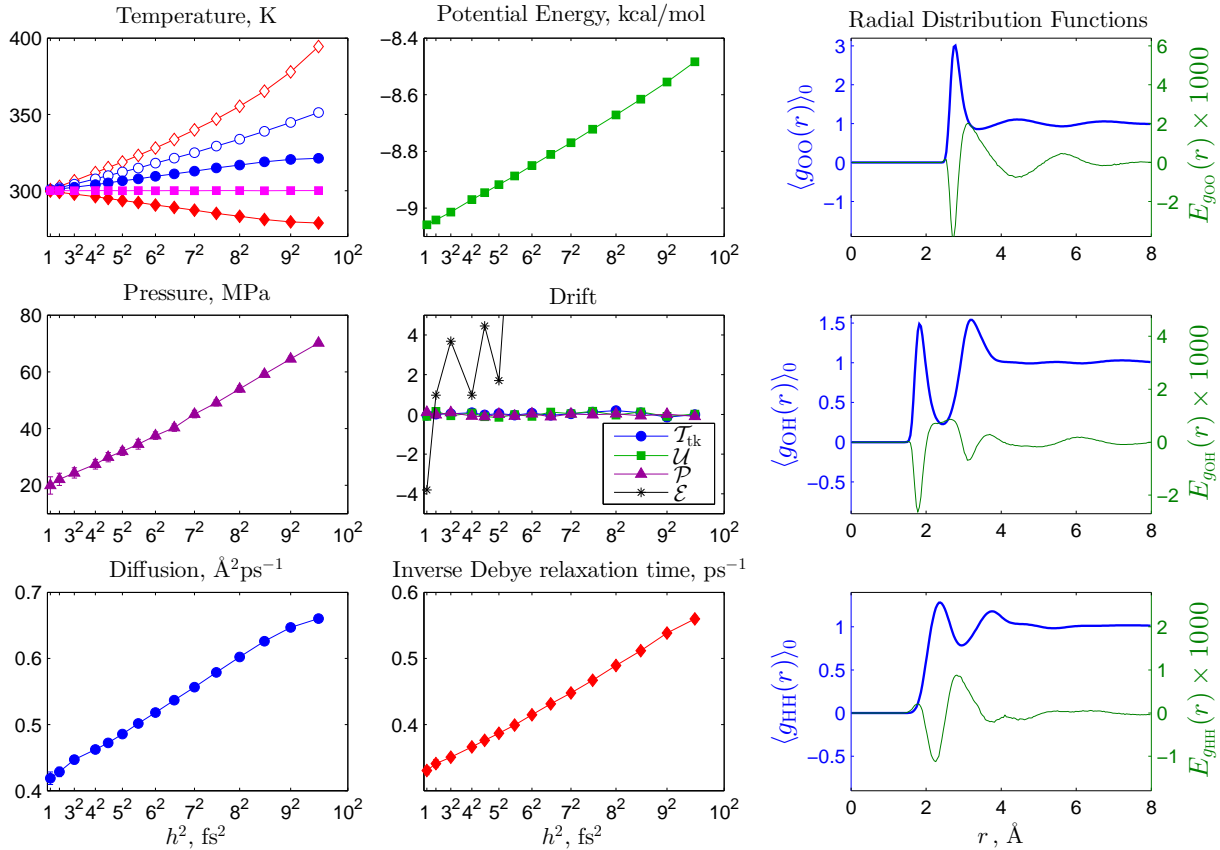}
\caption{\label{fig:nhi} (Color online) Results of simulations with the NH-I integrator.  Format of the plots is the same as in Figure~\ref{fig:nve}.  The RDF quantities $\langle g_{\alpha\beta}(r) \rangle_0$ and $E_{g_{\alpha\beta}}(r)$ were determined from the straight line least-squares fit to $\langle g_{\alpha\beta}(r) \rangle_h$ at each $r$ for $h \leq 7$\,fs.}
\end{figure}
\begin{table}[!tb]\caption{\label{tab:nhi}
Results of simulations with the NH-I integrator. Values for $\langle A \rangle_0$ and $E_{A}$, as defined in Eq.~(\ref{eq:h2err}), were determined from the straight line least-squares fit to $\langle A \rangle_h$ for $h \leq 7$\,fs.}
{\small
\begin{center}
\begin{tabular}{lcc}
\hline $A$ & $\langle A \rangle_0$ & $E_A$ \\\hline
$\mathcal{T}_\mathrm{tk}$,~K & $ 299.9(2)$ & $  0.261(6)$\\
$\mathcal{T}_\mathrm{rk}$,~K & $ 300.1(2)$ & $ -0.261(6)$\\
$\mathcal{T}_\mathrm{k}$,~K  & $ 300.00(9)$ & $ -0.000(3)$\\
$\mathcal{T}_\mathrm{tc}$,~K & $ 299.8(3)$ & $  0.505(11)$\\
$\mathcal{T}_\mathrm{rc}$,~K & $ 299.7(4)$ & $  0.809(10)$\\
$\mathcal{U}$,~kcal/mol  & $ -9.067(3)$ & $ 0.00604(7)$\\
$\mathcal{P}$,~MPa       & $  19.6(1.3)$ & $ 0.50(4)$\\
$\mathcal{D}$,~\AA$^2$ps$^{-1}$ & $  0.417(3)$ & $ 0.00282(9)$\\
\rld{$\tau^{-1}_\mathrm{D}$,~ps$^{-1}$}   & $  0.329(2)$ & $ 0.00240(6)$\\\hline
\end{tabular}\end{center}}\end{table}
\subsubsection{Implicit Nos\'e-Hoover (NH-I) Integrator}
It is not difficult to derive a Nos\'e-Hoover integrator which controls the temperature using the momentum at full steps.  One such integrator can be found in Refs.~\cite{FrenkelBook,Bond99} (see \ref{app:nhi} for details).  It is slightly more complicated than NH-E in that it requires solution of a cubic equation for $\xi^{n+1}$, which can be easily done using Newton's iterations (hence the word 'implicit' in the name).  The results are shown in Figure~\ref{fig:nhi} and Table~\ref{tab:nhi}.  Now the total kinetic temperature is maintained equal to the thermostat temperature at all step sizes.  At the same time, other temperatures display similar shift in discretization error compared to those for the NH-E integrator.  In fact, comparing results in Tables~\ref{tab:nhe} and \ref{tab:nhi} we see that $E_A$ for all temperatures are shifted by approximately 0.5\,K/fs$^2$.  The discretization errors in other quantities, such as potential energy, pressure, diffusion, RDFs, etc., are shifted as well, so that these errors are now much larger than for the NH-E integrator.   Since it is these quantities, rather than the temperature, that are usually of interest, the ability of the integrator to exactly control the kinetic temperature has, somewhat surprisingly, a detrimental effect on its overall performance.

\begin{figure}[!tb]
\includegraphics[width=16cm]{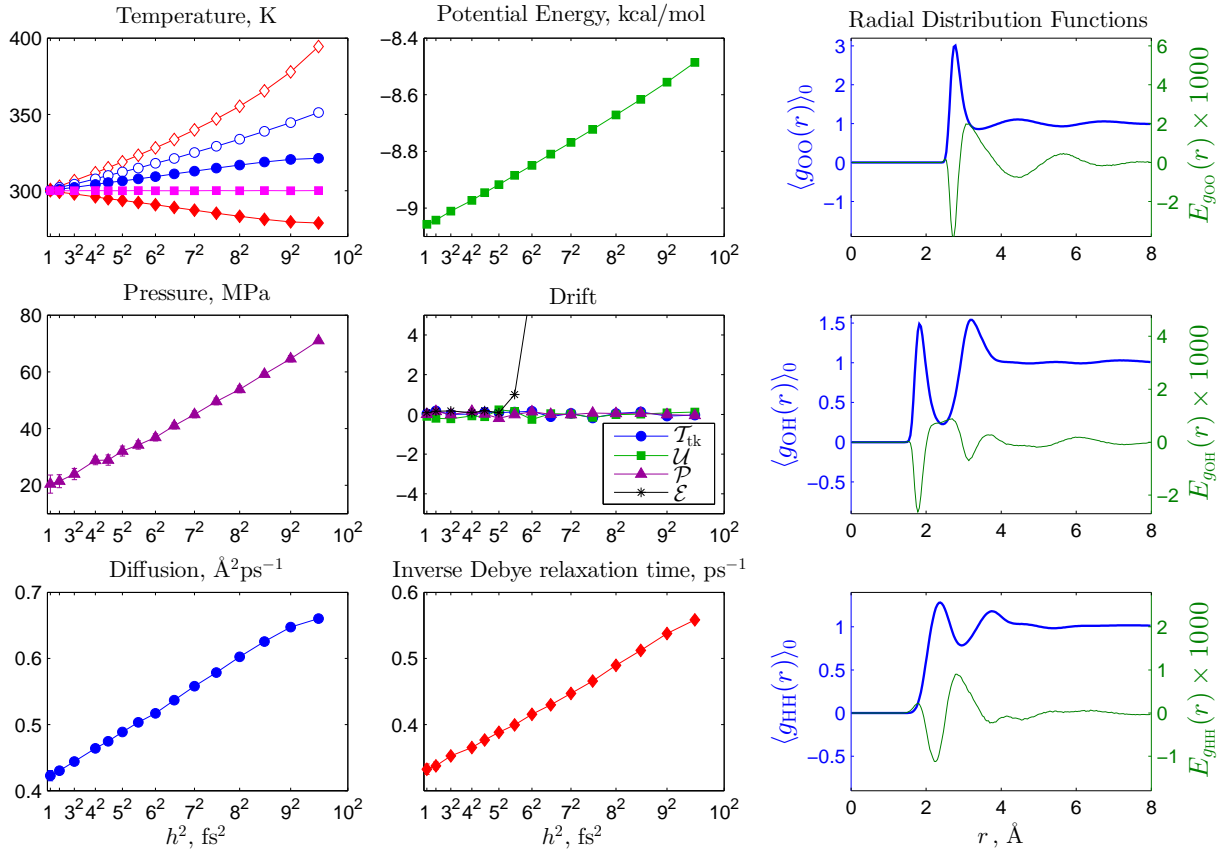}
\caption{\label{fig:nhmp} (Color online) Results of simulations with the NH-MP integrator.  Format of the plots is the same as in Figure~\ref{fig:nve}.  The RDF quantities $\langle g_{\alpha\beta}(r) \rangle_0$ and $E_{g_{\alpha\beta}}(r)$ were determined from the straight line least-squares fit to $\langle g_{\alpha\beta}(r) \rangle_h$ at each $r$ for $h \leq 7$\,fs.}
\end{figure}
\begin{table}[!tb]\caption{ \label{tab:nhmp}
Results of simulations with the NH-MP integrator. Values for $\langle A \rangle_0$ and $E_{A}$, as defined in Eq.~(\ref{eq:h2err}), were determined from the straight line least-squares fit to $\langle A \rangle_h$ for $h \leq 7$\,fs.}
{\small
\begin{center}
\begin{tabular}{lcc}
\hline $A$ & $\langle A \rangle_0$ & $E_A$ \\\hline
$\mathcal{T}_\mathrm{tk}$,~K & $ 299.9(3)$ & $  0.262(7)$\\
$\mathcal{T}_\mathrm{rk}$,~K & $ 300.1(3)$ & $ -0.262(7)$\\
$\mathcal{T}_\mathrm{k}$,~K  & $ 300.00(10)$ & $ -0.000(3)$\\
$\mathcal{T}_\mathrm{tc}$,~K & $ 299.9(3)$ & $  0.506(10)$\\
$\mathcal{T}_\mathrm{rc}$,~K & $ 299.7(4)$ & $  0.809(11)$\\
$\mathcal{U}$,~kcal/mol  & $ -9.067(3)$ & $ 0.00605(8)$\\
$\mathcal{P}$,~MPa       & $  19.5(1.4)$ & $ 0.50(4)$\\
$\mathcal{D}$,~\AA$^2$ps$^{-1}$ & $  0.419(3)$ & $ 0.00279(7)$\\
\rld{$\tau^{-1}_\mathrm{D}$,~ps$^{-1}$}   & $  0.329(3)$ & $ 0.00239(7)$\\\hline
\end{tabular}\end{center}}\end{table}
\subsubsection{Measure-Preserving Nos\'e-Hoover (NH-MP) Integrator}
Another Nos\'e-Hoover integrator we have tested has the property of preserving the invariant measure in the extended space~\cite{Tuckerman92}.  Its complexity is similar to that of NH-E, in that it is fully explicit and easy to code, while, similar to NH-I, it controls the total kinetic energy measured at full steps.  The results are shown in Figure~\ref{fig:nhmp} and Table~\ref{tab:nhmp}.  It turns out that these results are identical, within the statistical error, to those obtained with the NH-I integrator. The only observed difference between NH-I and NH-MP, which is probably due to the measure-preserving property of NH-MP, is that the drift of the extended energy is less pronounced than in the other two integrators and, as can be seen in Figure~\ref{fig:nhmp}, becomes significant at about 5.5\,fs step size.  However, just like with NH-E and NH-I, all other measured quantities remain stationary up to the stability threshold, so the improved conservation of the extended energy doesn't appear to be relevant to the overall performance of the integrator.

\begin{figure}[!tb]
\includegraphics[width=16cm]{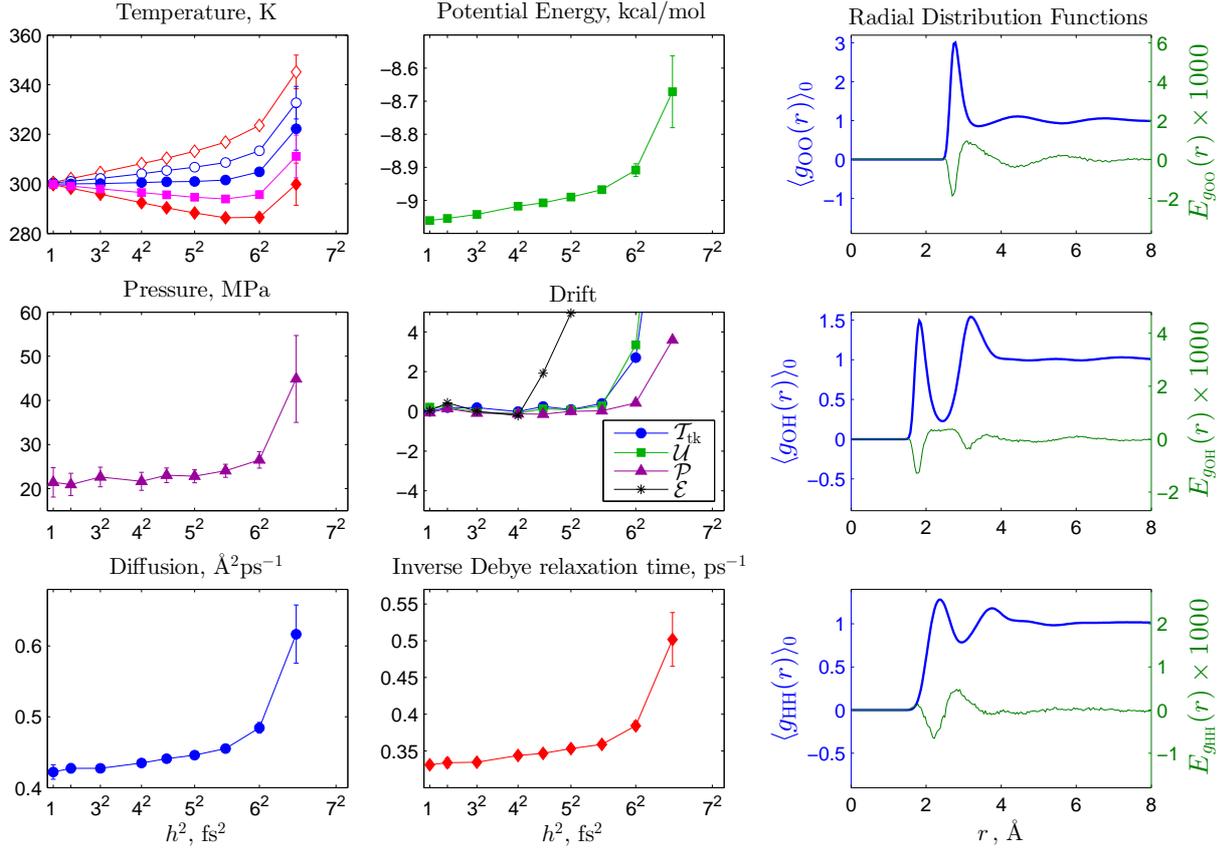}
\caption{\label{fig:np} (Color online) Results of simulations with the NP integrator.  Format of the plots is the same as in Figure~\ref{fig:nve}.  The RDF quantities $\langle g_{\alpha\beta}(r) \rangle_0$ and $E_{g_{\alpha\beta}}(r)$ were determined from the straight line least-squares fit to $\langle g_{\alpha\beta}(r) \rangle_h$ at each $r$ for $h \leq 5$\,fs.}
\end{figure}
\begin{table}[!tb]\caption{\label{tab:np}
Results of simulations with the NP integrator. Values for $\langle A \rangle_0$ and $E_{A}$, as defined in Eq.~(\ref{eq:h2err}), were determined from the straight line least-squares fit to $\langle A \rangle_h$ for $h \leq 5$\,fs.}
{\small
\begin{center}
\begin{tabular}{lcc}
\hline $A$ & $\langle A \rangle_0$ & $E_A$ \\\hline
$\mathcal{T}_\mathrm{tk}$,~K & $ 299.8(3)$ & $  0.046(18)$\\
$\mathcal{T}_\mathrm{rk}$,~K & $ 300.1(3)$ & $ -0.478(18)$\\
$\mathcal{T}_\mathrm{k}$,~K  & $ 299.95(14)$ & $ -0.216(7)$\\
$\mathcal{T}_\mathrm{tc}$,~K & $ 300.0(5)$ & $  0.26(3)$\\
$\mathcal{T}_\mathrm{rc}$,~K & $ 300.0(5)$ & $  0.52(3)$\\
$\mathcal{U}$,~kcal/mol  & $ -9.066(4)$ & $ 0.00297(19)$\\
$\mathcal{P}$,~MPa       & $  21.2(2.1)$ & $ 0.07(11)$\\
$\mathcal{D}$,~\AA$^2$ps$^{-1}$ & $  0.421(5)$ & $ 0.0010(3)$\\
\rld{$\tau^{-1}_\mathrm{D}$,~ps$^{-1}$}   & $  0.329(3)$ & $ 0.00090(16)$\\\hline
\end{tabular}\end{center}}\end{table}
\subsection{Nos\'e-Poincar\'e (NP) Integrator}
It is possible to modify the Nos\'e-Hoover dynamics in order to restore the Hamiltonian structure of the equations of motion in the extended space, allowing the construction of a Nos\'e-Poincar\'e integrator~\cite{Bond99}, which is symplectic.  The details of the integrator can be found in \ref{app:np}.  The results are shown in Figure~\ref{fig:np} and Table~\ref{tab:np}.  Unlike the Nos\'e-Hoover integrators considered above, the NP integrator does not have a stabilizing effect on the system dynamics, with the stability threshold at about $h = 7\,$fs, similar to that of the V-NSQ integrator.  The linear dependence of discretization errors on $h^2$ extends to about $h = 5\,$fs.  Also similar to the V-NSQ integrator, all measured quantities experience drift in simulations with larger step sizes.  The total kinetic temperature is not controlled exactly, with the value of $E_{\mathcal{T}_\mathrm{k}}$ between those for the NH-E and NH-I (NH-MP) thermostats.  Similarly, the values of $E_A$ for all measured quantities are approximately in the middle between those for NH-E and NH-I (NH-MP).

\subsection{Separate Nos\'e-Hoover thermostats}
In cases when the system consists of several types of degrees of freedom that evolve on significantly different timescales, the coupling of all degrees of freedom to a single thermostat may lead to inefficient equilibration due to a very slow energy flow (weak coupling) between different types of degrees of freedom.  In such cases it is often recommended that different types of degrees of freedom are coupled to separate Nos\'e-Hoover thermostats, thus providing independent temperature control for each type of degrees of freedom~\cite{Hunenberger05}.  For example, in the system of water molecules investigated in this work, the translational and rotational degrees of freedom could be coupled to two separate Nos\'e-Hoover thermostats.  The equations of motion for the momenta, Eq.~(\ref{eq:NHEM}), would be modified as follows:
\begin{align}\label{eq:NHEMsep}
  \dot{p}_i =& \;f_i - \xi p_i\,, \nonumber\\
  \dot{\pi}_i =& -\!\nabla_{q_i} K_\mathrm{rot}(q,\pi) + F_i - \zeta \pi_i\,,
\end{align}
where $\xi$ and $\zeta$ are scalar dynamic variables evolving according to equations
\begin{align}\label{eq:xi-zeta}
  \dot{\xi} =& \;\frac{1}{Q_\mathrm{tra}}[2K_\mathrm{tra}(p) - (3N-3)\kb T]\,,\nonumber\\
  \dot{\zeta} =& \;\frac{1}{Q_\mathrm{rot}}[2K_\mathrm{rot}(q,\pi) - 3N\kb T]
\end{align}
with initial conditions $\xi(0) = 0$ and $\zeta(0) = 0$.  The new dynamics preserves the extended energy in the form
\begin{equation}\label{eq:NHexensep}
  \mathcal{E} = H(\mathbf{r}, \mathbf{p}, \mathbf{q}, \bm{\pi}) + \tfrac{1}{2}Q_\mathrm{tra}\xi^2 + (3N-3)\kb T \eta + \tfrac{1}{2}Q_\mathrm{rot}\zeta^2 + 3N\kb T \nu\,,
\end{equation}
where $\dot{\eta} = \xi$, $\eta(0) = 0$, and $\dot{\nu} = \zeta$, $\nu(0) = 0$.

We have implemented separate Nos\'e-Hoover thermostats using three different integrators, NH-E, NH-I, and NH-MP, and in all cases observed undesirable side effects of the discretization error on the system evolution.  Even for small step sizes, where the extended energy did not exhibit any significant drift, we observed that the thermostat variables $\eta(t)$ and $\nu(t)$ drifted in opposite directions, indicating that the thermal energy was flowing through the system from one thermostat to the other.  So, rather than modelling the system in a thermal equilibrium, such simulation appeared to model a steady non-equilibrium process, with a steady flow of energy from one type of degrees of freedom to another.  With NH-E, the energy flowed from translational to rotational degrees of freedom, while with NH-I and NH-MP integrators it flowed in the opposite direction.

\begin{figure}[!tb]
\includegraphics[width=16cm]{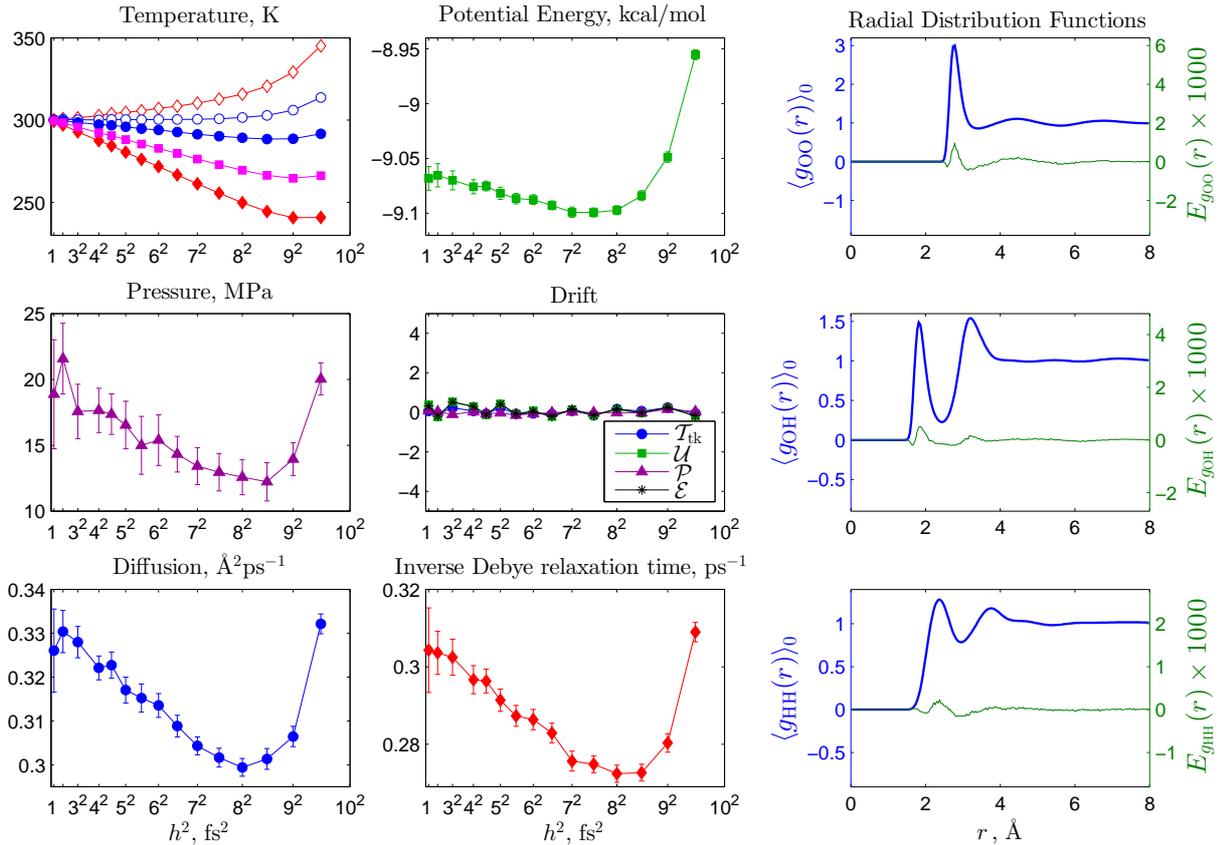}
\caption{\label{fig:lana} Results of simulations with the Langevin A integrator with $\gamma = 5\,$ps$^{-1}$ and $\Gamma = 10\,$ps$^{-1}$.  Format of the plots is the same as in Figure~\ref{fig:nve}.  The RDF quantities $\langle g_{\alpha\beta}(r) \rangle_0$ and $E_{g_{\alpha\beta}}(r)$ were determined from the straight line least-squares fit to $\langle g_{\alpha\beta}(r) \rangle_h$ at each $r$ for $h \leq 6$\,fs.}
\end{figure}
Another undesirable consequence of using separate Nos\'e-Hoover thermostats is that, with the NH-E integrator, the total linear momentum of the system, which is set equal to zero (to within machine precision) at the start of the simulation, grows exponentially with time.  This indicates that the zero total momentum state, which is conserved by the exact solutions of the Nos\'e-Hoover equations (\ref{eq:NHEMsep})-(\ref{eq:xi-zeta}), becomes linearly unstable when the equations are solved with the NH-E integrator.  The instability is stronger for larger step sizes, but is present at all step sizes.  With the other two integrators, we have observed instabilities related to the overall rotational dynamics of the system, but we have not investigated their precise origin.

\begin{figure}[!tb]
\includegraphics[width=16cm]{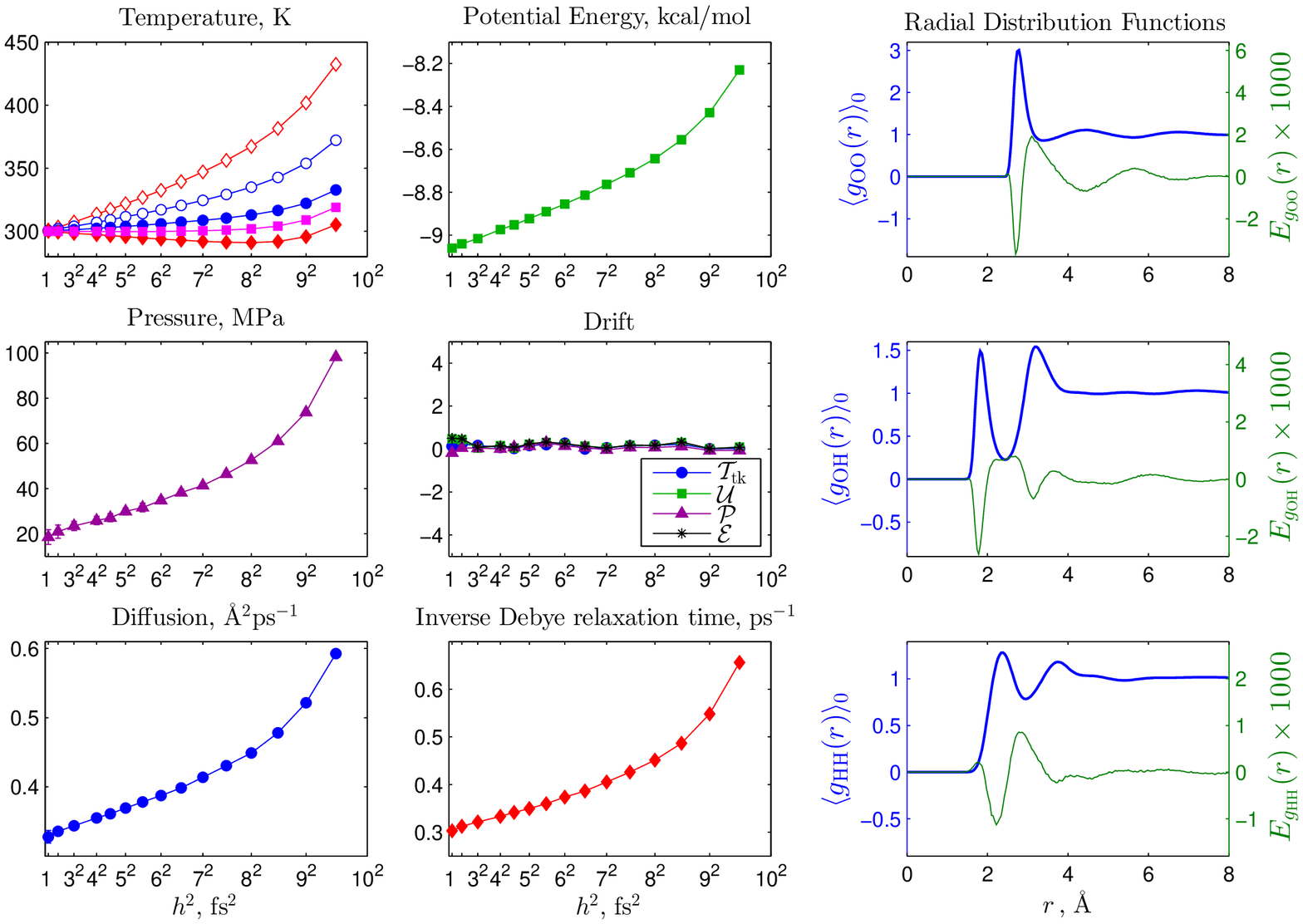}
\caption{\label{fig:lanb} Results of simulations with the Langevin B integrator with $\gamma = 5\,$ps$^{-1}$ and $\Gamma = 10\,$ps$^{-1}$.  Format of the plots is the same as in Figure~\ref{fig:nve}.  The RDF quantities $\langle g_{\alpha\beta}(r) \rangle_0$ and $E_{g_{\alpha\beta}}(r)$ were determined from the straight line least-squares fit to $\langle g_{\alpha\beta}(r) \rangle_h$ at each $r$ for $h \leq 6$\,fs.}
\end{figure}
\subsection{Langevin Thermostat (Lan-A and Lan-B)}
Another approach to simulating the canonical ensemble is to couple the system to a stochastic thermostat described by the Langevin equation, which is well known for the translational degrees of freedom and has been recently proposed for the rotational ones~\cite{Davidchack09}.  The combined Langevin equations for both translational and rotational dynamics can be written in the form
\begin{align}\label{eq:LanEM}
  d r_i =& \;\frac{p_i}{m} dt\,, \nonumber\\
  d q_i =& \;\nabla_{\pi_i} K_\mathrm{rot}(\mathbf{q},\bm{\pi}) dt\,, \nonumber\\
  d p_i =& \;f_i dt - \gamma p_i dt + \sqrt{2m\kb T\gamma}dw_i(t)\,, \nonumber\\
  d \pi_i =& -\!\nabla_{q_i} K_\mathrm{rot}(\mathbf{q},\bm{\pi}) dt + F_i dt - \Gamma J(q_i) \pi_i dt + \sqrt{2M\kb T\Gamma}dW_i(t)\,,
\end{align}
where $\gamma$ and $\Gamma$, measured in the units of inverse time, control the coupling of the translational and rotational degrees of freedom to the thermostat, $M^{-1} = \frac{1}{4}\sum_{k=1}^3 I_k^{-1}$, $J(q) = \frac{1}{4}M\sum_{k=1}^3 I_k^{-1} \mathbf{S}_k q [\mathbf{S}_k q]\tran$, and $w_i$ and $W_i$ denote the standard Wiener processes in 3 and 4 dimensions, respectively.  Matrices $\mathbf{S}_k$ are defined in \ref{app:app}. 

\begin{table}[!tb]\caption{\label{tab:lan01-02} Results of simulations with Langevin A and Langevin B integrators with weak thermostat coupling: $\gamma = 1\,$ps$^{-1}$ and $\Gamma = 2\,$ps$^{-1}$.  Values for $\langle A \rangle_0$ and $E_{A}$, as defined in Eq.~(\ref{eq:h2err}), were determined from the straight line least-squares fit to $\langle A \rangle_h$ for $h \leq 6$\,fs.}
{\small
\begin{center}
\begin{tabular}{lccccc}
\hline & \multicolumn{2}{c}{Langevin A} &{\ }& \multicolumn{2}{c}{Langevin B}\\\cline{2-3}\cline{5-6}
$A$ & $\langle A \rangle_0$ & $E_A$ &{\ }& $\langle A \rangle_0$ & $E_A$ \\\hline
$\mathcal{T}_\mathrm{tk}$,~K & $ 300.4(6)$ & $ -0.22(3)$ &{\ }& $ 300.2(7)$ & $  0.22(3)$\\
$\mathcal{T}_\mathrm{rk}$,~K & $ 300.3(8)$ & $ -0.76(3)$ &{\ }& $ 299.8(8)$ & $ -0.24(3)$\\
$\mathcal{T}_\mathrm{k}$,~K  & $ 300.3(7)$ & $ -0.49(3)$ &{\ }& $ 300.0(7)$ & $ -0.01(3)$\\
$\mathcal{T}_\mathrm{tc}$,~K & $ 300.3(6)$ & $ -0.01(3)$ &{\ }& $ 299.9(8)$ & $  0.48(3)$\\
$\mathcal{T}_\mathrm{rc}$,~K & $ 300.0(7)$ & $  0.24(3)$ &{\ }& $ 299.5(8)$ & $  0.80(3)$\\
$\mathcal{U}$,~kcal/mol  & $ -9.062(10)$ & $-0.0008(4)$ &{\ }& $ -9.068(10)$ & $ 0.0058(4)$\\
$\mathcal{P}$,~MPa       & $  19.7(1.9)$ & $-0.15(8)$ &{\ }& $  20.5(2.3)$ & $ 0.42(9)$\\
$\mathcal{D}$,~\AA$^2$ps$^{-1}$ & $  0.408(5)$ & $-0.00083(19)$ &{\ }& $  0.405(7)$ & $ 0.0023(3)$\\
\rld{$\tau^{-1}_\mathrm{D}$,~ps$^{-1}$}   & $  0.329(5)$ & $-0.00079(17)$ &{\ }& $  0.322(5)$ & $ 0.00215(19)$\\\hline
\end{tabular}\end{center}}\end{table}

\begin{table}[!tb]\caption{\label{tab:lan05-10} Results of simulations with Langevin A and Langevin B integrators with moderate thermostat coupling: $\gamma = 5\,$ps$^{-1}$ and $\Gamma = 10\,$ps$^{-1}$.  Values for $\langle A \rangle_0$ and $E_{A}$, as defined in Eq.~(\ref{eq:h2err}), were determined from the straight line least-squares fit to $\langle A \rangle_h$ for $h \leq 6$\,fs.}
{\small
\begin{center}
\begin{tabular}{lccccc}
\hline & \multicolumn{2}{c}{Langevin A} &{\ }& \multicolumn{2}{c}{Langevin B}\\\cline{2-3}\cline{5-6}
$A$ & $\langle A \rangle_0$ & $E_A$ &{\ }& $\langle A \rangle_0$ & $E_A$ \\\hline
$\mathcal{T}_\mathrm{tk}$,~K & $ 300.2(4)$ & $ -0.175(17)$ &{\ }& $ 299.9(4)$ & $  0.161(17)$\\
$\mathcal{T}_\mathrm{rk}$,~K & $ 300.1(4)$ & $ -0.787(17)$ &{\ }& $ 300.1(5)$ & $ -0.174(18)$\\
$\mathcal{T}_\mathrm{k}$,~K  & $ 300.1(4)$ & $ -0.481(16)$ &{\ }& $ 300.0(4)$ & $ -0.006(14)$\\
$\mathcal{T}_\mathrm{tc}$,~K & $ 300.3(5)$ & $  0.00(2)$ &{\ }& $ 299.9(5)$ & $  0.468(19)$\\
$\mathcal{T}_\mathrm{rc}$,~K & $ 299.9(5)$ & $  0.19(2)$ &{\ }& $ 299.6(5)$ & $  0.899(19)$\\
$\mathcal{U}$,~kcal/mol  & $ -9.065(6)$ & $-0.0007(3)$ &{\ }& $ -9.067(6)$ & $ 0.0058(2)$\\
$\mathcal{P}$,~MPa       & $  20.0(2.0)$ & $-0.14(9)$ &{\ }& $  18.8(1.8)$ & $ 0.44(7)$\\
$\mathcal{D}$,~\AA$^2$ps$^{-1}$ & $  0.330(4)$ & $-0.00046(14)$ &{\ }& $  0.328(4)$ & $ 0.00166(14)$\\
\rld{$\tau^{-1}_\mathrm{D}$,~ps$^{-1}$}   & $  0.306(4)$ & $-0.00056(17)$ &{\ }& $  0.303(3)$ & $ 0.00192(11)$\\\hline
\end{tabular}\end{center}}\end{table}
The Langevin dynamics, as could be expected, also has a stabilizing effect on the system dynamics.  Stabilization is stronger with larger values of $\gamma$ and/or $\Gamma$: with $\gamma = 1\,$ps$^{-1}$ and $\Gamma = 0$ or $\gamma = 0$ and $\Gamma = 2\,$ps$^{-1}$ the system is stable up to step sizes of about 8.5\,fs, while the maximum stability threshold of about 10\,fs is reached when $\gamma + \Gamma$ is larger than about 8\,ps$^{-1}$.  The results with $\gamma = 5\,$ps$^{-1}$ and $\Gamma = 10\,$ps$^{-1}$ are shown in Figure~\ref{fig:lana} for the simulation with Lan-A and Figure~\ref{fig:lanb} for the simulation with Lan-B.  The linear dependence of $\langle A \rangle_h$ on $h^2$ extends to about 6\,fs.  There in no drift in the measured quantities.  (For the Langevin thermostats, we define $\mathcal{E} = K_\mathrm{tra} + K_\mathrm{rot} + U$.)  Broadly, the discretization errors look similar to those with the Nos\'e-Hoover thermostats.  The only significant difference is that, with the Langevin thermostats, the dynamic quantities, $\langle \mathcal{D} \rangle_h$ and $\langle \tau^{-1}_\mathrm{D} \rangle_h$, do not converge to correct values as $h \to 0$.  This is due to the well know fact that Langevin dynamics distorts the physical time of a dynamical process.  The distortion increases with increasing values of the thermostat coupling parameters $\gamma$ and $\Gamma$.  This can be clearly observed when comparing results presented in Tables~\ref{tab:lan01-02}-\ref{tab:lan00-10}.  For the weak coupling, shown in Table~\ref{tab:lan01-02}, the results for $\langle \mathcal{D} \rangle_0$ and $\langle \tau^{-1}_\mathrm{D}\rangle_0$ are close to those obtained with the Nos\'e-Hoover thermostats, while with increased coupling both quantities become significantly smaller.  It is also interesting to note that, as can be seen in Table~\ref{tab:lan00-10}, when only rotational degrees of freedom are coupled to the thermostat (i.e., when $\gamma = 0$), the dynamical quantities remain unaffected (although we do see some deviation when the coupling becomes very strong, as in the simulation with $\gamma = 0$ and $\Gamma = 50\,$ps$^{-1}$).

We have investigated discretization errors and their dependence on $\gamma$ and $\Gamma$ for two different integrators: Langevin A (Lan-A) and Langevin B (Lan-B)~\cite{Davidchack09}.  Both integrators are second order (in the weak sense), quasi-symplectic, and exactly preserve the constraint $|q_i| = 1$.  
With $\gamma = \Gamma = 0$, both Lan-A and Lan-B reduce to the V-NSQ integrator.  Since the integrators are not time reversible, the dependence of $\langle A \rangle_h$ on $h$ is expected to be described by Eq.~(\ref{eq:h2err}) with $p = 3$.
\begin{table}[!tb]\caption{\label{tab:lan25-50} Results of simulations with Langevin A and Langevin B integrators with strong thermostat coupling: $\gamma = 25\,$ps$^{-1}$ and $\Gamma = 50\,$ps$^{-1}$.  Values for $\langle A \rangle_0$ and $E_{A}$, as defined in Eq.~(\ref{eq:h2err}), were determined from the straight line least-squares fit to $\langle A \rangle_h$ for $h \leq 6$\,fs.}
{\small
\begin{center}
\begin{tabular}{lccccc}
\hline & \multicolumn{2}{c}{Langevin A} &{\ }& \multicolumn{2}{c}{Langevin B}\\\cline{2-3}\cline{5-6}
$A$ & $\langle A \rangle_0$ & $E_A$ &{\ }& $\langle A \rangle_0$ & $E_A$ \\\hline
$\mathcal{T}_\mathrm{tk}$,~K & $ 300.0(3)$ & $ -0.126(10)$ &{\ }& $ 299.90(18)$ & $  0.077(6)$\\
$\mathcal{T}_\mathrm{rk}$,~K & $ 300.1(3)$ & $ -0.850(11)$ &{\ }& $ 300.1(2)$ & $ -0.028(7)$\\
$\mathcal{T}_\mathrm{k}$,~K  & $ 300.01(19)$ & $ -0.488(8)$ &{\ }& $ 300.02(15)$ & $  0.025(5)$\\
$\mathcal{T}_\mathrm{tc}$,~K & $ 300.0(5)$ & $  0.02(2)$ &{\ }& $ 299.6(5)$ & $  0.497(14)$\\
$\mathcal{T}_\mathrm{rc}$,~K & $ 299.9(4)$ & $  0.098(17)$ &{\ }& $ 299.7(4)$ & $  1.191(11)$\\
$\mathcal{U}$,~kcal/mol  & $ -9.067(4)$ & $-0.00055(17)$ &{\ }& $ -9.071(4)$ & $ 0.00629(11)$\\
$\mathcal{P}$,~MPa       & $  20.0(2.3)$ & $-0.07(9)$ &{\ }& $19.5(1.9)$ & $ 0.34(6)$\\
$\mathcal{D}$,~\AA$^2$ps$^{-1}$ & $  0.1913(16)$ & $-0.00016(6)$ &{\ }& $0.1907(8)$ & $ 0.00075(2)$\\
\rld{$\tau^{-1}_\mathrm{D}$,~ps$^{-1}$}  & $0.2343(15)$ & $-0.00035(6)$ &{\ }& $0.2305(19)$ & $ 0.00145(6)$\\\hline
\end{tabular}\end{center}}\end{table}
\begin{table}[!tb]\caption{\label{tab:lan05-00} Results of simulations with Langevin A and Langevin B integrators with coupling of the thermostat only to translational degrees of freedom: $\gamma = 5\,$ps$^{-1}$ and $\Gamma = 0$.  Values for $\langle A \rangle_0$ and $E_{A}$, as defined in Eq.~(\ref{eq:h2err}), were determined from the straight line least-squares fit to $\langle A \rangle_h$ for $h \leq 6$\,fs.}
{\small
\begin{center}
\begin{tabular}{lccccc}
\hline & \multicolumn{2}{c}{Langevin A} &{\ }& \multicolumn{2}{c}{Langevin B}\\\cline{2-3}\cline{5-6}
$A$ & $\langle A \rangle_0$ & $E_A$ &{\ }& $\langle A \rangle_0$ & $E_A$ \\\hline
$\mathcal{T}_\mathrm{tk}$,~K & $ 299.9(4)$ & $ -0.085(18)$ &{\ }& $ 300.0(4)$ & $ -0.000(18)$\\
$\mathcal{T}_\mathrm{rk}$,~K & $ 300.3(7)$ & $ -0.63(3)$ &{\ }& $ 299.9(6)$ & $ -0.49(3)$\\
$\mathcal{T}_\mathrm{k}$,~K  & $ 300.1(5)$ & $ -0.36(2)$ &{\ }& $ 299.9(4)$ & $ -0.246(18)$\\
$\mathcal{T}_\mathrm{tc}$,~K & $ 300.0(5)$ & $  0.13(2)$ &{\ }& $ 299.9(5)$ & $  0.25(2)$\\
$\mathcal{T}_\mathrm{rc}$,~K & $ 300.1(6)$ & $  0.38(3)$ &{\ }& $ 299.6(6)$ & $  0.52(2)$\\
$\mathcal{U}$,~kcal/mol  & $ -9.066(7)$ & $ 0.0011(3)$ &{\ }& $ -9.067(7)$ & $ 0.0026(3)$\\
$\mathcal{P}$,~MPa       & $  19.9(1.7)$ & $ 0.01(7)$ &{\ }& $  19.4(1.9)$ & $ 0.14(8)$\\
$\mathcal{D}$,~\AA$^2$ps$^{-1}$ & $  0.332(2)$ & $ 0.00004(8)$ &{\ }& $  0.330(3)$ & $ 0.00055(12)$\\
\rld{$\tau^{-1}_\mathrm{D}$,~ps$^{-1}$}  & $  0.307(2)$ & $ 0.00005(8)$ &{\ }& $  0.306(3)$ & $ 0.00062(13)$\\\hline
\end{tabular}\end{center}}\end{table}
\begin{table}[!tb]\caption{\label{tab:lan00-10} Results of simulations with Langevin A and Langevin B integrators with coupling of the thermostat only to rotational degrees of freedom: $\gamma = 0$ and $\Gamma = 10$ps$^{-1}$.  Values for $\langle A \rangle_0$ and $E_{A}$, as defined in Eq.~(\ref{eq:h2err}), were determined from the straight line least-squares fit to $\langle A \rangle_h$ for $h \leq 6$\,fs.}
{\small
\begin{center}
\begin{tabular}{lccccc}
\hline & \multicolumn{2}{c}{Langevin A} &{\ }& \multicolumn{2}{c}{Langevin B}\\\cline{2-3}\cline{5-6}
$A$ & $\langle A \rangle_0$ & $E_A$ &{\ }& $\langle A \rangle_0$ & $E_A$ \\\hline
$\mathcal{T}_\mathrm{tk}$,~K & $ 300.0(8)$ & $ -0.40(3)$ &{\ }& $ 299.5(6)$ & $  0.60(3)$\\
$\mathcal{T}_\mathrm{rk}$,~K & $ 300.2(5)$ & $ -0.92(2)$ &{\ }& $ 299.6(6)$ & $  0.08(2)$\\
$\mathcal{T}_\mathrm{k}$,~K  & $ 300.1(6)$ & $ -0.66(3)$ &{\ }& $ 299.6(5)$ & $  0.34(2)$\\
$\mathcal{T}_\mathrm{tc}$,~K & $ 300.2(7)$ & $ -0.20(3)$ &{\ }& $ 299.5(6)$ & $  0.85(3)$\\
$\mathcal{T}_\mathrm{rc}$,~K & $ 300.0(6)$ & $  0.06(2)$ &{\ }& $ 299.4(6)$ & $  1.16(2)$\\
$\mathcal{U}$,~kcal/mol  & $ -9.065(10)$ & $-0.0033(4)$ &{\ }& $ -9.071(8)$ & $ 0.0104(4)$\\
$\mathcal{P}$,~MPa       & $  19.6(1.9)$ & $-0.37(8)$ &{\ }& $  18.8(1.7)$ & $ 1.00(7)$\\
$\mathcal{D}$,~\AA$^2$ps$^{-1}$ & $  0.417(6)$ & $-0.0021(3)$ &{\ }& $  0.411(4)$ & $ 0.00542(16)$\\
\rld{$\tau^{-1}_\mathrm{D}$,~ps$^{-1}$}   & $  0.330(5)$ & $-0.0018(2)$ &{\ }& $  0.324(5)$ & $ 0.0046(2)$\\\hline
\end{tabular}\end{center}}\end{table}

Unlike the Nos\'e-Hoover and Nos\'e-Poincar\'e integrators, where the discretization errors appear to be independent of the thermostat parameter $Q$ in a wide range of parameters values, the errors for the Langevin integrators depend on $\gamma$ and $\Gamma$.  We have varied the two parameters independently and obtained results for $\gamma = 0, 1, 5, 25\,$ps$^{-1}$, and $\Gamma = 0, 2, 10, 50\,$ps$^{-1}$. (Note that the values $\gamma = 5\,$ps$^{-1}$ and $\Gamma = 10\,$ps$^{-1}$ were determined in Ref.~\cite{Davidchack09} to be close to optimal with respect to the speed of system equilibration for TIP4P liquid water at 270\,K.)  We will not report here the results of all 15 simulations for each thermostat, but rather report selected results and describe the trends of the dependence of the results on the thermostat parameters.

Contrasting Lan-A and Lan-B we see that Lan-B is better at maintaining the correct values of the kinetic temperatures, but other measured quantities have larger discretization errors than with Lan-A.  In this respect, the difference between Lan-A and Lan-B is similar to that between the NH-E and NH-I (or NH-MP) integrators.

\section{Correcting Discretization Errors} \label{sec:correct}
Accounting for discretization errors and correcting their influence on the results can be done using backward error analysis, as discussed in the Introduction.  The work is underway to implement this approach for various systems and thermostats\cite{BondPrivate}.  Until such tools are available, we present below a couple of practical approaches that might be considered, which allow for correcting or removing the discretization errors from the measured quantities.

\subsection{Richardson Extrapolation} \label{sec:richardson}
To correct for the discretization error in the measured quantities, we can run simulations with several different step sizes and then extrapolate the result to the limit of zero step size.  This procedure is known in numerical analysis as Richardson extrapolation~\cite{Joyce71}.  Below we present the analysis of the Richardson extrapolation procedure when two simulations are performed with different step sizes and the extrapolation allows to remove the $h^2$ discretization error term from the results obtained with second order numerical integrators.

As we have seen in this work, if the equations of motion are solved using a second order numerical integrator, then the estimated average obtained from a simulation run with step size $h$ has the following step size dependence:
\[\langle A \rangle_h = \langle A \rangle_0 + h^2 E_A + \mathcal{O}(h^p)\,, \]
where $p = 4$ if the integrator is time reversible and $p = 3$ otherwise. If a second simulation run is performed with step size $sh$, $0 < s < 1$, then we can also write
\[ \langle A \rangle_{sh} = \langle A \rangle_0 + (sh)^2 E_A + \mathcal{O}(h^p)\,. \]
The two expressions can be combined to yield a higher order estimator for $\langle A \rangle_0$:
\begin{equation}\label{eq:Rich}
  \langle A \rangle_0 = \frac{\langle A \rangle_{sh} - s^2\langle A \rangle_h}{1-s^2} + O(h^p)\,.
\end{equation}
This expression is a particular case of a more general Richardson extrapolation formula, which is often used in numerical analysis in order to accelerate the rate of convergence of numerical methods~\cite{Joyce71}.

To determine the optimal choice for $s$ and for the relative duration of the two runs, we need to minimize the estimated statistical error of $\langle A \rangle_0$ in Eq.~(\ref{eq:Rich}).  In a simulation run of $L$ steps with step size $h$, the average of $A(t)$ along an equilibrium trajectory is given by
\[  \langle A \rangle_h = \frac{1}{L}\sum_{l=1}^L A(lh)\,. \]
The statistical error of this quantity can be estimated from the variance of the sample $A(lh)$ and is inversely proportional to the total simulation time $t_\mathrm{max} = hL$:
\begin{equation}\label{eq:varAh}
  \mathrm{var}\;\langle A \rangle_h = \frac{\tau_A}{t_\mathrm{max}} \mathrm{var}\;A(lh)\,,
\end{equation}
assuming that the system dynamics is chaotic and $t_\mathrm{max}$ is much larger than the characteristic correlation time $\tau_A$.  The correlation time is generally not known {\em a priori}, but its value can be estimated from the calculation of block averages~\cite{AllenBook} of the sample $A(lh)$.

\begin{figure}[tbp]
\begin{center} \includegraphics[width=8cm]{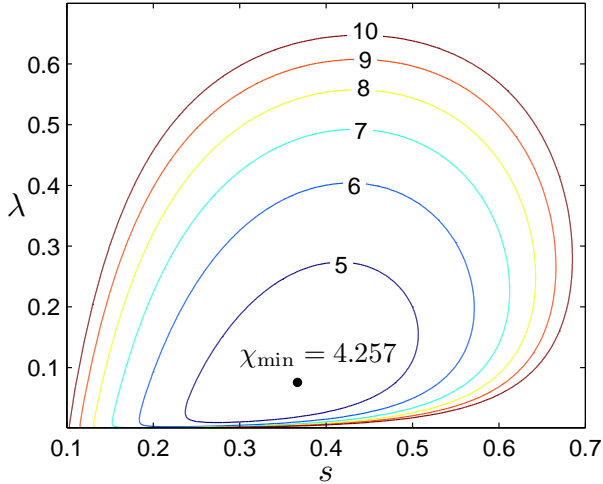} \end{center}
\caption{(Color online) Contour plot of the function $\chi(s,\lambda)$ defined in Eq.~(\ref{eq:chi}).}
\label{fig:rich} \end{figure}
Let us assume that the duration of the simulation run with the step size $h$ is $\lambda L$ steps, $0 < \lambda < 1$, while that of the run with the step size $sh$ is $(1-\lambda)L$.  Then, using the property of the variance $\mathrm{var}(a X + b Y) = a^2 \mathrm{var}\;X + b^2 \mathrm{var}\;Y$ for two independent measurements $X$ and $Y$ and two constants $a$ and $b$, we can express the variance of $\langle A \rangle_0$ as follows:
\[ \mathrm{var}\;\langle A \rangle_0 = \chi(s,\lambda) \frac{\tau_A}{t_\mathrm{max}} \mathrm{var}\;A(lh) \]
where
\begin{equation}\label{eq:chi}
  \chi(s,\lambda) = \frac{s^{-1}(1-\lambda)^{-1} + s^4\lambda^{-1}}{(1-s^2)^2}\,,
\end{equation}
and we assume that $\mathrm{var}\;A(lh) = \mathrm{var}\;A(lsh)$, i.e. the variance of the sample is independent of the time step.  (This assumption should be true for all measured quantities except those that are conserved by the dynamics.)  As can be seen in Figure~\ref{fig:rich}, for $0 < s < 1$ and $0 < \lambda < 1$, the function $\chi(s, \lambda)$ has a single minimum at $s_\mathrm{min} \approx 0.367$ and $\lambda_\mathrm{min} \approx 0.0755$ with the value $\chi(s_\mathrm{min}, \mu_\mathrm{min}) \approx 4.257$.

These results demonstrate that, in order for Eq.~(\ref{eq:Rich}) to yield an estimate with statistical error comparable to that of a single $L$ step run, it is necessary to be able to run the simulation with the step size $h$ that is about four times larger than that of the single run.  Such an estimate will be accurate if the value of the $O(h^p)$ term in Eq.~(\ref{eq:Rich}) remains negligible at this $h$ compared to the statistical error.  From the results reported above we could see that in the simulations of TIP4P water with the Nos\'e-Hoover thermostats the higher order terms could be neglected for step sizes up to about 7\,fs.  Since majority of simulations of the rigid water models are performed with steps sizes of about 0.5 or 1\,fs, Richardson extrapolation would allow to obtain in this case higher precision results with the same computational effort or, alternatively, would allow to reduce the computational effort. 

Finally, note that $0.92L$ steps are carried out with the time step $0.367h$ for the time duration of $0.34hL$, while only $0.08L$ steps with the time step $h$, for the time duration of $0.08hL$, are required to find the optimal correction for the order $h^2$ discretization error.   It is important that $L$ is large enough so that $0.08hL$ is much larger than the correlation time $\tau_A$ for any quantity of interest.

\subsection{Weighted Thermostating} \label{sec:weighted}
The weighted thermostating  approach to removing discretization errors is based on the observation that the values of the discretization errors $E_A$ depend on the way the system is coupled to a thermostat.  In the case of the Nos\'e-Hoover thermostat coupled to the system with translational and rotational degrees of freedom this can be exploited by introducing different weights for translational and rotational kinetic energies in Eq.~(\ref{eq:xi}) for the thermostat variable:
\begin{equation}\label{eq:xiwei}
  \dot{\xi} = \frac{1}{Q}\big\{w\big[2K_\mathrm{tra}(p) - (3N-3)\kb T\big] + (2-w)\big[2K_\mathrm{rot}(q,\pi) - 3N\kb T\big]\big\}\,.
\end{equation}
For $w = 1$ this reduces to Eq.~(\ref{eq:xi}), while for $w = 2$ ($w = 0$) the Nos\'e-Hoover thermostat is coupled only to translational (rotational) degrees of freedom.  Since now $E_A$ are expected to be functions of $w$, we hope to find such value of $w = w_A$ that, for a particular measured quantity of interest $A$, the leading term in the discretization error is zero, $E_A(w_A) = 0$.

\begin{figure}[tbp]
\includegraphics[width=16cm]{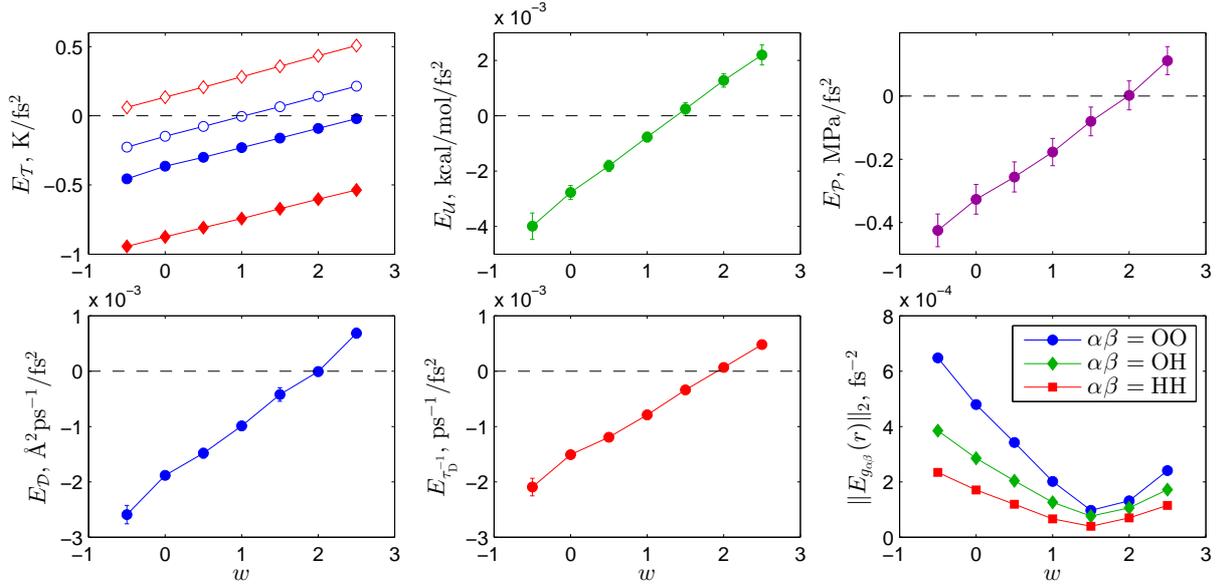}
\caption{(Color online) Dependence of discretization errors on $w$ within the weighted thermostating approach using NH-E integrator.  The symbols for different measured temperatures are the same as in Figure~\ref{fig:nve}.}
\label{fig:wei} \end{figure}

\begin{figure}[tbp]
\includegraphics[width=16cm]{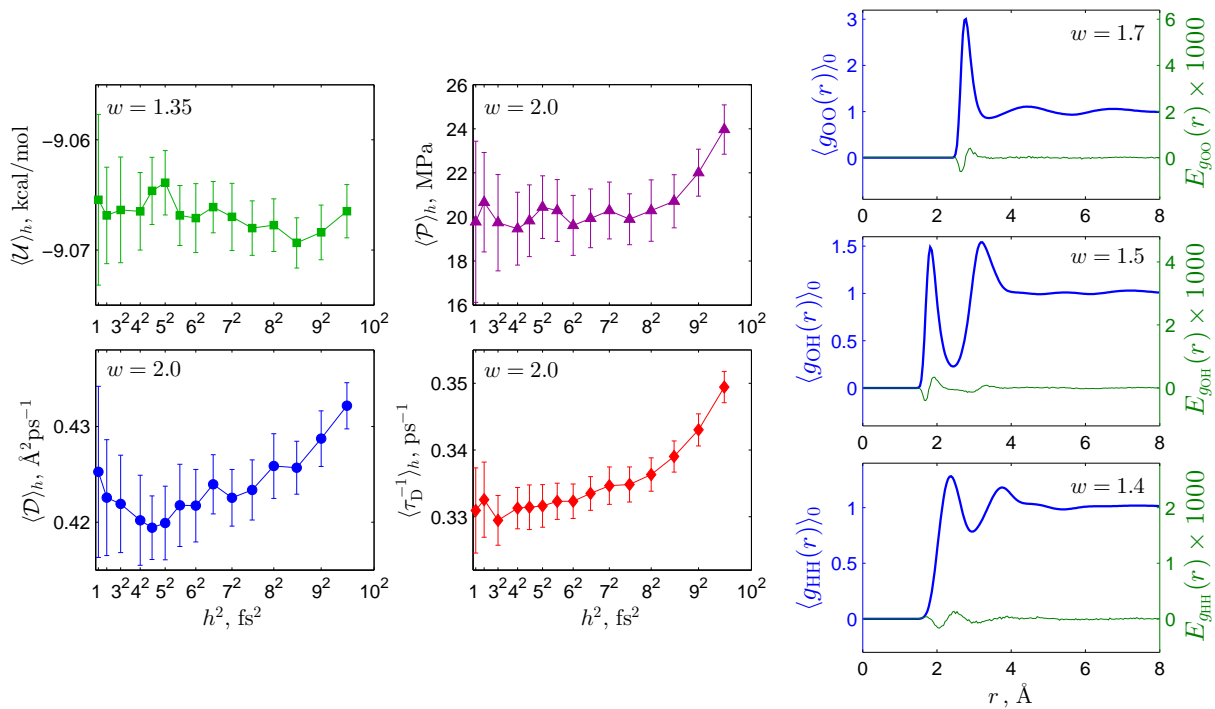}
\caption{(Color online) Results for optimal choices of $w$ within the weighted thermostating approach using NH-E integrator.}
\label{fig:weiopt} \end{figure}

We have implemented the weighed thermostat with NH-E, NH-I, and NH-MP.  As an example, in Figure~\ref{fig:wei} we show the results for NH-E.  Here, as well as with the other two integrators, we see the linear dependence of the discretization errors on $w$ for temperatures, potential energy, pressure, diffusion, and Debye relaxation time.  Therefore, it is a relatively simple task to find a value of $w$ for which the discretization error is equal to zero.  Thus we find that $w_\mathcal{U} = 1.35$ and $w_\mathcal{P} = w_\mathcal{D} = w_{\tau^{-1}_\mathrm{D}} = 2.0$.   For the simulations with NH-I and NH-MP, the weights are $w_\mathcal{U} = 2.7$ and $w_\mathcal{P} = w_\mathcal{D} = w_{\tau^{-1}_\mathrm{D}} = 2.37$, so that the rotational kinetic energy has a negative weight.  For the errors in the RDFs we also find the linear dependence of $E_{g_{\alpha\beta}}(r)$ on $w$ for each value of $r$.   However, since for different $r$ we need somewhat different values of $w$ in order to eliminate the $h^2$ term from Eq.~(\ref{eq:h2rdf}), it is impractical to adopt such an approach.  Instead, we can try to find the value of $w$ for which the norm of $E_{g_{\alpha\beta}}(r)$ is minimized.  In Figure~\ref{fig:wei} we plot the $L^2$ norms of $E_{g_{\alpha\beta}}(r)$, defined as
\[
  \| E_{g_{\alpha\beta}}(r) \|_2 = \left(\frac{1}{r_\mathrm{max}}\int_0^{r_\mathrm{max}} E^2_{g_{\alpha\beta}}(r) dr \right)^{1/2}
\]
with $r_\mathrm{max} = 8\,$\AA. For different RDFs, the smallest error is observed for similar, but not exactly the same, values of $w$: it is 1.7 for $g_\mathrm{OO}$, 1.5 for $g_\mathrm{OH}$, and 1.4 for $g_\mathrm{HH}$.  The results for the weighted thermostating with the NH-E integrator and the optimal choice of the weights are shown in Figure~\ref{fig:weiopt} and Table~\ref{tab:weiopt}.

\begin{table}[tbp]\caption{\label{tab:weiopt}
Results of simulations with the weighted NH-E integrator. Values for $\langle A \rangle_0$ and $E_{A}$, as defined in Eq.~(\ref{eq:h2err}), were determined from the straight line least-squares fit to $\langle A \rangle_h$ for $h \leq 7$\,fs.}
{\small
\begin{center}
\begin{tabular}{lccc}
\hline $A$ & $w_A$ & $\langle A \rangle_0$ & $E_A$ \\\hline
$\mathcal{U}$,~kcal/mol  & 1.35 & $ -9.066(3)$ & $ -0.00001(10)$\\
$\mathcal{P}$,~MPa       & 2.0 & $  20.0(1.5)$ & $ 0.00(5)$\\
$\mathcal{D}$,~\AA$^2$ps$^{-1}$ & 2.0 & $  0.422(3)$ & $ -0.00001(8)$\\
\rld{$\tau^{-1}_\mathrm{D}$,~ps$^{-1}$}  & 2.0 & $  0.320(2)$ & $ 0.00007(6)$\\\hline
\end{tabular}\end{center}}\end{table}

Even if the system is homogeneous with only one type of degrees of freedom, the weighted thermostating approach can be used by introducing weighted coupling to kinetic and configurational temperatures.  The Nos\'e-Hoover thermostat coupled to the configurational temperature has already been proposed~\cite{Travis08}, and it should be straightforward to combine it with the kinetic temperature thermostat.

\section{Summary} \label{sec:summary}
We have measured discretization errors in static, dynamic, and structural quantities in the molecular dynamics simulation of a system of TIP4P liquid water coupled to various Nos\'e-Hoover and Langevin thermostats.  Based on the analysis of the obtained results, we list the following main observations:
\begin{itemize}
\item All measured quantities exhibit dependence on the step size as described by Eq.~(\ref{eq:h2err}), as expected from the backward error analysis for second order numerical integrators.  The linear dependence of discretization errors on $h^2$ extends up to about 70\% of the stability threshold of the integrators.
\item Both Nos\'e-Hoover and Langevin thermostats have a stabilizing effect on the numerical integration of the equations of motion, increasing the stability threshold from about 7\,fs for the isolated system (integrated with V-NSQ) to about 10\,fs for the Nos\'e-Hoover thermostats (for a wide range of physically reasonable values of the thermostat parameter $\tau_\mathrm{NH}$) and 8.5-10\,fs for the Langevin thermostats (increasing with the coupling strength).  The No\'e-Poincar\'e thermostat does not stabilize the dynamics.
\item The results with the Nos\'e-Hoover and Nos\'e-Poincar\'e thermostats are independent of $\tau_\mathrm{NH}$ in a wide range of physically reasonable parameter values: $\tau_\mathrm{NH} = 20$-500\,fs.  The results with the Langevin thermostats depend on the coupling parameters $\gamma$ and $\Gamma$.
\item Different methods of measuring the system temperature (kinetic and configurational, rotational and translational), exhibit different discretization errors, although they all consistently converge to the thermostat temperature in the limit $h \to 0$.  The errors are larger in magnitude for the rotational temperatures and, for all the integrators studied here, are ordered as follows: $E_{\mathcal{T}_\mathrm{rk}} <  E_{\mathcal{T}_\mathrm{tk}} <  E_{\mathcal{T}_\mathrm{tc}} <  E_{\mathcal{T}_\mathrm{rc}}$.
\item All measured quantities in the simulations with Nos\'e-Hoover and Nos\'e-Poincar\'e thermostats consistently converge to the same values in the limit $h \to 0$.  For the Langevin thermostats, the static and structural quantities also converge to the same values, while the dynamical quantities, $\langle \mathcal{D} \rangle_0$ and $\langle \tau^{-1}_\mathrm{D}\rangle_0$, deviate from their correct values, with the deviation increasing for larger values of the coupling parameters $\gamma$ and $\Gamma$.
\item Even though the Nos\'e-Hoover integrators are constructed to control the total kinetic temperature of the system, the measured $\mathcal{T}_\mathrm{k}$ for the NH-E integrator deviates from the thermostat temperature.  This is due to the fact that the temperature in the NH-E integrator is controlled at half steps and there is a difference between the average kinetic energy of the system measured at half steps and at full steps.  By contrast, the NH-I and NH-MP integrators precisely control $\mathcal{T}_\mathrm{k}$ at all step sizes up to the stability threshold.  However, the relationship between the discretization error for the temperature and that for other quantities (potential energy, pressure, diffusion, Debye relaxation time, RDFs) is such that precise control of the total kinetic temperature results in large discretization errors in other quantities.  In this respect the less precise control of the NH-E integrator is preferable.
\item Moderate instability of the integrators manifests itself in a gradual drift of the total (extended) energy of the system, which is conserved by the exact dynamics.  While for the V-NSQ and NP integrators this drift is also present in other measured quantities, for the Nos\'e-Hoover integrators the drift is localized within the thermostat variables and thus is not indicative on the non-stationary evolution within the system itself; all measured quantities of the system coupled to a Nos\'e-Hoover thermostat remain stationary up to the stability threshold.
\item To thermostat systems with several different types of degrees of freedom that evolve on significantly different timescales, it is often recommended that separate thermostats are used to control the temperature of different types of degrees of freedom.  We have implemented such an approach in our system by coupling the translational and rotational degrees of freedom to two separate Nos\'e-Hoover thermostats and observed undesirable side effects of the discretization errors on the system evolution.  In particular, even when simulating a well equilibrated system, the thermostats induced a steady flow of energy from one type of degrees of freedom to the other.  Therefore, we recommend that the coupling of the system to several thermostats should be avoided.
\end{itemize}

We have proposed two possible approaches for taking the discretization errors into account in order to obtain accurate measurements in simulations with large time steps.  The first approach is based on running the simulation with two different step sizes and then using Richardson extrapolation to eliminate the leading term in the discretization error.  Our analysis shows that, given the total computational budget of $L$ steps, the most precise measurements can be obtained when the first simulation is performed with step size $h$ for $0.08L$ steps and the second simulation is performed with step size $0.367h$ for $0.92L$ steps.  Our study suggests that $h$ can be as large as 70\% of the stability threshold of the integrator, which for the system investigated here is about 7\,fs.

The second approach is based on a weighted coupling of the Nos\'e-Hoover thermostat to different types of degrees of freedom.  For each quantity of interest, $A$, it is possible to find the value of the weight parameter, $w_A$, such that the leading discretization error term is zero, $E_A(w_A) = 0$.  The obvious drawback of this approach is that, in the simulation with a chosen weight parameter, the discretization error can be removed for only one quantity, unless the optimal weights for several quantities of interest coincide.

Finally, we would like to point out that larger step sizes should be used with caution when simulating systems close to phase or other transition regions, since it is possible that the discretization errors shift the system state across the thermodynamic phase boundary into a phase which is different from the one modeled by the exact dynamics.  In this case the simple linear dependence of the errors on $h^2$ may break down at smaller step sizes than those demonstrated in the present study.

\section{Acknowledgments} \label{sec:acknw}
I would like to thank Brian Laird, Stephen Bond, Ben Leimkuhler, Mark Tuckerman, and Glen Martyna for fruitful discussions and helpful remarks.  The computations have been carried out on the University of Leicester HPC cluster purchased through the HEFCE Science Research Investment Fund.

\appendix
\section{Numerical Integrator Schemes} \label{app:app}
The rotational equations of motion are integrated using the symplectic method NO\_SQUISH proposed by Miller {\em et al.}\cite{Miller02}.  The method is based on rewriting the rotational kinetic energy of a molecule in the form
\[\tfrac{1}{8}\pi\tran \mathbf{S}(q)\mathbf{I}^{-1}\mathbf{S}\tran(q)\pi 
= \sum_{k=1}^3 \frac{(\pi\tran\mathbf{S}_k q)^2}{8I_k}\,, \]
where $I_1 = I_{xx}$, $I_2 = I_{yy}$, $I_3 = I_{zz}$, and the constant matrices $\mathbf{S}_k$ are defined as follows:
\begin{align*}
  \mathbf{S}_1 q =& (-q^1, q^0, q^3, -q^2)\tran,\nonumber\\
  \mathbf{S}_2 q =& (-q^2, -q^3, q^0, q^1)\tran,\nonumber\\
  \mathbf{S}_3 q =& (-q^3, q^2, -q^1, q^0)\tran.
\end{align*}
This allows to introduce a second order integrator for free rotations $(q(t),\pi(t)) = \mathcal{R}^t(q(0),\pi(0))$:
\[ \mathcal{R}^t = \mathcal{R}_1^{t/2} \circ \mathcal{R}_2^{t/2} \circ \mathcal{R}_3^{t} \circ \mathcal{R}_2^{t/2} \circ \mathcal{R}_1^{t/2}\,,\]
where $(\bar{q},\bar{\pi}) = \mathcal{R}_k^{t}(q,\pi)$ is defined as follows:
\begin{align*}
  \bar{q} =& \cos(\zeta_k t) q + \sin(\zeta_k t)\mathbf{S}_k q\,,\\
  \bar{\pi} =& \cos(\zeta_k t)\pi + \sin(\zeta_k t)\mathbf{S}_k \pi\,,
\end{align*}
with
\[\zeta_k = \frac{1}{4I_k}\pi\tran\mathbf{S}_k q\,. \]
This integrator is symplectic, time reversible, and exactly preserves the constraint $|q| = 1$.

The following numerical integrators were investigated:

\subsection{Constant energy Verlet-NO\_SQUISH (V-NSQ) integrator} \label{app:nve}
This integrator combines velocity-Verlet algorithm for translational and NO\_SQUISH algorithm~\cite{Miller02} for rotational dynamics.  The combined integrator is second order, symplectic, and time reversible.
\begin{align}
  p_i^{n+\frac{1}{2}} =& \;p_i^n + \tfrac{h}{2} f_{i}^n\,,\nonumber\\
  \pi_i^{n+\frac{1}{2}} =& \;\pi_i^n + \tfrac{h}{2} F_{i}^n\,,\nonumber\\
  r_i^{n+1} =& \;r_i^n + \tfrac{h}{m} p_i^{n+\frac{1}{2}}\,,\nonumber\\
  (q_i^{n+1}, &\bar{\pi}_i^{n+\frac{1}{2}}) = \;\mathcal{R}^h(q_i^{n}, \pi_i^{n+\frac{1}{2}})\,,\nonumber\\
  p_i^{n+1} =& \;p_i^{n+\frac{1}{2}} + \tfrac{h}{2} f_{i}^{n+1}\,,\nonumber\\
  \pi_i^{n+1} =& \;\bar{\pi}_i^{n+\frac{1}{2}} + \tfrac{h}{2} F_{i}^{n+1}\,,
\end{align}
where $f_{i}^n = -\nabla_{r_i} U(\mathbf{r}^n, \mathbf{q}^n)$ is the translational force acting on the center of mass of molecule $i$ and $F_{i}^n = -\tilde{\nabla}_{q_i} U(\mathbf{r}^n, \mathbf{q}^n)$ is the rotational force, which can be expressed through the torque acting on molecule $i$~\cite{Miller02}.

\subsection{Explicit Nos\'e-Hoover (NH-E) integrator} \label{app:nhe}
This integrator is based on the one described in Refs.~\cite{Holian90,Bond99}, which has been extended to include thermostating of rotational degrees of freedom.
\begin{align}
  p_i^{n+\frac{1}{2}} =& \left(p_i^n + \tfrac{h}{2} f_{i}^n\right)/\left(1 + \tfrac{h}{2}\xi^n\right), \nonumber\\
  \pi_i^{n+\frac{1}{2}} =& \left(\pi_i^n + \tfrac{h}{2} F_{i}^n\right)/\left(1 + \tfrac{h}{2}\xi^n\right),\nonumber\\
  r_i^{n+1} =& \;r_i^n + \tfrac{h}{m} p_i^{n+\frac{1}{2}}\,,\nonumber\\
  (q_i^{n+1}, &\bar{\pi}_i^{n+\frac{1}{2}}) = \;\mathcal{R}^h(q_i^{n}, \pi_i^{n+\frac{1}{2}}),\nonumber\\
  \xi^{n+1} =&\;\xi^n + \tfrac{h}{Q}\big[2K_\mathrm{tra}\big(\mathbf{p}^{n+\frac{1}{2}}\big)
              + 2K_\mathrm{rot}\big(\mathbf{q}^{n+1},\bar{\bm{\pi}}^{n+\frac{1}{2}}\big) - N_\mathrm{df}\kb T\big],\nonumber\\
  \eta^{n+1} =& \;\eta^n + \tfrac{h}{2} (\xi^n + \xi^{n+1}),\nonumber\\
  p_i^{n+1} =& \;p_i^{n+\frac{1}{2}}\left(1 - \tfrac{h}{2}\xi^{n+1}\right) + \tfrac{h}{2}f_{i}^{n+1},\nonumber\\
  \pi_i^{n+1} =& \;\bar{\pi}_i^{n+\frac{1}{2}}\left(1 - \tfrac{h}{2}\xi^{n+1}\right) + \tfrac{h}{2} F_{i}^{n+1},
\end{align}
where $N_\mathrm{df} = 6N-3$ is the number of thermostated degrees of freedom (taking into account the constraints of the total momentum conservation).  In the construction of the rotational part of the integrator we use the fact that the free rotation map $\mathcal{R}^t$ does not change the rotational kinetic energy of the system, i.e. $K_\mathrm{rot}\circ\mathcal{R}^t = K_\mathrm{rot}$, and so $K_\mathrm{rot}(\mathbf{q}^{n+1}, \bar{\bm{\pi}}^{n+\frac{1}{2}}) = K_\mathrm{rot}(\mathbf{q}^{n}, \bm{\pi}^{n+\frac{1}{2}})$.  The initial conditions for the thermostat variables are $\xi^0 = 0$, $\eta^0 = 0$.   Note that here, as well as in other Nos\'e-Hoover integrators described below, the thermostat variable $\eta$ does not appear in the equations for other variables, so it does not need to be integrated unless one wants to monitor the extended energy of the system defined in Eq.~(\ref{eq:NHexen}).

\subsection{Implicit Nos\'e-Hoover (NH-I) integrator} \label{app:nhi}
This is a straightforward extension of the integrator presented in Ref.~\cite{Bond99}, which includes rotational dynamics.
\begin{align}
  \xi^{n+\frac{1}{2}} =& \;\xi^n + \tfrac{h}{2Q}\big[2K_\mathrm{tra}\big(\mathbf{p}^{n}\big)
  + 2K_\mathrm{rot}\big(\mathbf{q}^{n},\bm{\pi}^{n}\big) - N_\mathrm{df}\kb T\big]\nonumber\\
  \eta^{n+1} =& \;\eta^n + h\xi^{n+\frac{1}{2}}\nonumber\\
  p_i^{n+\frac{1}{2}} =& \;p_i^n\left(1 - \tfrac{h}{2}\xi^n\right) + \tfrac{h}{2} f_{i}^n\nonumber\\
  \pi_i^{n+\frac{1}{2}} =& \;\pi_i^n\left(1 - \tfrac{h}{2}\xi^n\right) + \tfrac{h}{2} F_{i}^n\nonumber\\
  r_i^{n+1} =& \;r_i^n + \tfrac{h}{m} p_i^{n+\frac{1}{2}}\nonumber\\
  (q_i^{n+1}, &\bar{\pi}_i^{n+\frac{1}{2}}) = \;\mathcal{R}^h(q_i^{n}, \pi_i^{n+\frac{1}{2}})\nonumber\\
  p_i^{n+1} =& \;\big(p_i^{n+\frac{1}{2}} + \tfrac{h}{2}f_{i}^{n+1}\big)/\big(1 + \tfrac{h}{2}\xi^{n+1}\big)\nonumber\\
  \pi_i^{n+1} =& \;\big(\bar{\pi}_i^{n+\frac{1}{2}} + \tfrac{h}{2} F_{i}^{n+1}\big)/\big(1 + \tfrac{h}{2}\xi^{n+1}\big)\nonumber\\
  \xi^{n+1} =& \;\xi^{n+\frac{1}{2}} + \tfrac{h}{2Q}\big[2K_\mathrm{tra}\big(\mathbf{p}^{n+1}\big)
  + 2K_\mathrm{rot}\big(\mathbf{q}^{n+1},\bm{\pi}^{n+1}\big) - N_\mathrm{df}\kb T\big]
\end{align}
Note that the equations for $p_i^{n+1}$, $\pi_i^{n+1}$, and $\xi^{n+1}$ are coupled and must be solved together.  We solve these equations by substituting the expressions for $p_i^{n+1}$ and $\pi_i^{n+1}$ into the equation for $\xi^{n+1}$:
\begin{align*}
  \xi^{n+1} =& \;\xi^{n+\frac{1}{2}} + \tfrac{h}{2Q}\big\{\big[2K_\mathrm{tra}\big(\mathbf{p}^{n+\frac{1}{2}}+\tfrac{h}{2}\mathbf{f}^{n+1}\big)\\
  +& \;2K_\mathrm{rot}\big(\mathbf{q}^{n+1},\bar{\bm{\pi}}^{n+\frac{1}{2}}+\tfrac{h}{2}\mathbf{F}^{n+1}\big)\big]\big(1 + \tfrac{h}{2}\xi^{n+1}\big)^{-2}
  - N_\mathrm{df}\kb T\big\}
\end{align*}
and then applying the Newton-Raphson iteration
\[
  \xi_{j+1}^{n+1} = \xi_j^{n+1} - \frac{\xi_j^{n+1} - \xi^{n+\frac{1}{2}} - N(\xi_j^{n+1})}{1 - N^\prime_\xi(\xi_j^{n+1})}
\]
starting with
\[ \xi_0^{n+1} = \xi^{n+\frac{1}{2}} + N(\xi^{n+\frac{1}{2}}), \]
where
\begin{align*}
  N(\xi) =& \;\tfrac{h}{2Q}\big\{\big[2K_\mathrm{tra}\big(\mathbf{p}^{n+\frac{1}{2}}+\tfrac{h}{2}\mathbf{f}^{n+1}\big)\\
  +& \;2K_\mathrm{rot}\big(\mathbf{q}^{n+1},\bar{\bm{\pi}}^{n+\frac{1}{2}}+\tfrac{h}{2}\mathbf{F}^{n+1}\big)\big]\big(1 + \tfrac{h}{2}\xi\big)^{-2}
  - N_\mathrm{df}\kb T\big\}
\end{align*}
and
\begin{align*}
  N^\prime_\xi(\xi) =& \;-\tfrac{h^2}{2Q}\big[2K_\mathrm{tra}\big(\mathbf{p}^{n+\frac{1}{2}}+\tfrac{h}{2}\mathbf{f}^{n+1}\big)
  + 2K_\mathrm{rot}\big(\mathbf{q}^{n+1},\bar{\bm{\pi}}^{n+\frac{1}{2}}+\tfrac{h}{2}\mathbf{F}^{n+1}\big)\big]\big(1 + \tfrac{h}{2}\xi\big)^{-3}
\end{align*}
For the system modeled in this study, three iterations are sufficient to achieve double-precision accuracy even for the largest attempted step size of $h = 10\,$fs.

\subsection{Measure-Preserving Nos\'e-Hoover (NH-MP) integrator} \label{app:nhmp}
This is a straightforward extension of the integrator proposed in Ref.~\cite{Tuckerman92}.
\begin{align}
  \xi^{n+\frac{1}{2}} =& \;\xi^n + \tfrac{h}{2Q}\big[2K_\mathrm{tra}\big(\mathbf{p}^{n}\big)
  + 2K_\mathrm{rot}\big(\mathbf{q}^{n},\bm{\pi}^{n}\big) - N_\mathrm{df}\kb T\big]\nonumber\\
  \eta^{n+1} =& \;\eta^n + h\xi^{n+\frac{1}{2}}\nonumber\\
  p_i^{n+\frac{1}{2}} =& \;p_i^n\mathrm{e}^{-\frac{h}{2}\xi^{n+\frac{1}{2}}}
   + \tfrac{h}{2} f_{i}^n\nonumber\\
  \pi_i^{n+\frac{1}{2}} =& \;\pi_i^n\mathrm{e}^{-\frac{h}{2}\xi^{n+\frac{1}{2}}}
   + \tfrac{h}{2} F_{i}^n\nonumber\\
  r_i^{n+1} =& \;r_i^n + \tfrac{h}{m} p_i^{n+\frac{1}{2}}\hspace{-3cm}\nonumber\\
  (q_i^{n+1}, &\bar{\pi}_i^{n+\frac{1}{2}}) = \;\mathcal{R}^h(q_i^{n}, \pi_i^{n+\frac{1}{2}})\nonumber\\
  p_i^{n+1} =& \;\big(p_i^{n+\frac{1}{2}} + \tfrac{h}{2}f_{i}^{n+1}\big)\mathrm{e}^{-\frac{h}{2}\xi^{n+\frac{1}{2}}}\nonumber\\
  \pi_i^{n+1} =& \;\big(\bar{\pi}_i^{n+\frac{1}{2}} + \tfrac{h}{2} F_{i}^{n+1}\big)\mathrm{e}^{-\frac{h}{2}\xi^{n+\frac{1}{2}}}\nonumber\\
  \xi^{n+1} =& \;\xi^{n+\frac{1}{2}} + \tfrac{h}{2Q}\big[2K_\mathrm{tra}\big(\mathbf{p}^{n+1}\big)
  + 2K_\mathrm{rot}\big(\mathbf{q}^{n+1},\bm{\pi}^{n+1}\big) - N_\mathrm{df}\kb T\big]
\end{align}

\subsection{Nos\'e-Poincar\'e (NP) integrator} \label{app:np}
This is a straightforward extension of the integrator proposed in Ref.~\cite{Bond99}.
\begin{align}
  \tilde{p}_i^{n+\frac{1}{2}} =& \;\tilde{p}_i^n + \tfrac{h}{2} s^n f_{i}^n\nonumber\\
  \tilde{\pi}_i^{n+\frac{1}{2}} =& \;\tilde{\pi}_i^n + \tfrac{h}{2} s^n F_{i}^n\nonumber\\
  \eta^{n+\frac{1}{2}} =& \;\eta^n + \tfrac{h}{2}\big[K_\mathrm{tra}(\tilde{\mathbf{p}}^{n+\frac{1}{2}}/s^n)
  + K_\mathrm{rot}(\mathbf{q}^n, \tilde{\bm{\pi}}^{n+\frac{1}{2}}/s^n)\nonumber\\
  -& \;U(\mathbf{r}^n, \mathbf{q}^n) - (\eta^{n+\frac{1}{2}})^2/(2Q) - N_\mathrm{df}\kb T (1 + \ln s^n) + H_0 \big]\nonumber\\
  \tilde{h} =& \;\frac{2Qh}{(2Q + h\eta^{n+\frac{1}{2}})s^n}\nonumber\\
  s^{n+1} =& \;\frac{2Q + h\eta^{n+\frac{1}{2}}}{2Q - h\eta^{n+\frac{1}{2}}}s^n\nonumber\\
  r_i^{n+1} =& \;r_i^n + \tfrac{\tilde{h}}{m} \tilde{p}_i^{n+\frac{1}{2}}\nonumber\\
  \big(q_i^{n+1}, & \;\bar{\tilde{\pi}}_i^{n+\frac{1}{2}}\big) = \mathcal{R}^{\tilde{h}}\big(q_i^{n}, \tilde{\pi}_i^{n+\frac{1}{2}}\big)\nonumber\\
  \eta^{n+1} =& \;\eta^{n+\frac{1}{2}} + \tfrac{h}{2}\big[K_\mathrm{tra}(\tilde{\mathbf{p}}^{n+\frac{1}{2}}/s^{n+1})
  + K_\mathrm{rot}(\mathbf{q}^{n+1}, \bar{\tilde{\bm{\pi}}}^{n+\frac{1}{2}}/s^{n+1})\nonumber\\
  -& \;U(\mathbf{r}^{n+1}, \mathbf{q}^{n+1})\nonumber - (\eta^{n+\frac{1}{2}})^2/(2Q) - N_\mathrm{df}\kb T (1 + \ln s^{n+1}) + H_0 \big]\nonumber\\
  \tilde{p}_i^{n+1} =& \;\tilde{p}_i^{n+\frac{1}{2}} + \tfrac{h}{2}s^{n+1} f_{i}^{n+1}\nonumber\\
  \tilde{\pi}_i^{n+1} =& \;\bar{\tilde{\pi}}_i^{n+\frac{1}{2}} + \tfrac{h}{2} s^{n+1} F_{i}^{n+1}
\end{align}
This symplectic integrator is based on the Hamiltonian
\begin{equation} \label{eq:HamNP}
  H_\mathrm{NP}(\mathbf{r},\tilde{\mathbf{p}},\mathbf{q},\tilde{\bm{\pi}},s,\eta) = s\big[H(\mathbf{r},\tilde{\mathbf{p}}/s,\mathbf{q},\tilde{\bm{\pi}}/s)
   + \eta^2/(2Q) + N_\mathrm{df}\kb T\ln s - H_0\big]
\end{equation}
where $\tilde{\mathbf{p}} = s\mathbf{p}$ and $\tilde{\bm{\pi}} = s\bm{\pi}$ are the scaled momentum variables and $H_0 = H(\mathbf{r}(0), \mathbf{p}(0), \mathbf{q}(0), \bm{\pi}(0))$.  The initial conditions for the thermostat variables are $s^0 = 1$ and $\eta^0 = 0$.

This is an explicit method since the equation for $\eta^{n+\frac{1}{2}}$ is a quadratic equation which can be written as follows
\[ \frac{h}{4Q}\big(\eta^{n+\frac{1}{2}}\big)^2 + \eta^{n+\frac{1}{2}} + C = 0\,, \]
and the solution expressed in the form
\[ \eta^{n+\frac{1}{2}} = -2C/\big(1 + \sqrt{1 - hC/Q}\,\big)\,, \]
where
\begin{align*}
  C =& \;\tfrac{h}{2}\big[N_\mathrm{df}\kb T (1 + \ln s^n) - K_\mathrm{tra}(\tilde{\mathbf{p}}^{n+\frac{1}{2}}/s^n)\\
  -& \;K_\mathrm{rot}(\mathbf{q}^n, \tilde{\bm{\pi}}^{n+\frac{1}{2}}/s^n) + U(\mathbf{r}^n, \mathbf{q}^n) - H_0 \big] - \eta^n\,.
\end{align*}

\bibliographystyle{elsarticle-num}
\bibliography{mypubs,molsim,water}

\begin{thebibliography}{10}
\expandafter\ifx\csname url\endcsname\relax
  \def\url#1{\texttt{#1}}\fi
\expandafter\ifx\csname urlprefix\endcsname\relax\def\urlprefix{URL }\fi
\expandafter\ifx\csname href\endcsname\relax
  \def\href#1#2{#2} \def\path#1{#1}\fi

\bibitem{Groot97}
R.~D. Groot, P.~B. Warren, Dissipative particle dynamics: Bridging the gap
  between atomistic and mesoscopic simulation, J. Chem. Phys. 107~(11) (1997)
  4423--4435.

\bibitem{Jakobsen05}
A.~F. Jakobsen, O.~G. Mouritsen, G.~Besold, Artifacts in dynamical simulations
  of coarse-grained model lipid bilayers, J. Chem. Phys. 122~(20) (2005)
  204901.

\bibitem{Allen06}
M.~P. Allen, Configurational temperature in membrane simulations using
  dissipative particle dynamics, J. Phys. Chem. B 2006, 110, 3823-3830 110~(8)
  (2006) 3823--3830.

\bibitem{Hairer94}
E.~Hairer, Backward analysis of numerical integrators and symplectic methods,
  Ann. Numer. Math. 1 (1994) 107--132.

\bibitem{Benettin94}
G.~Benettin, A.~Giorgilli, On the {H}amiltonian interpolation of near to the
  identity symplectic mappings with application to symplectic integration
  algorithms, J. Statist. Phys. 74~(5-6) (1994) 1117–--1143.

\bibitem{HairerBookIII}
E.~Hairer, C.~Lubich, G.~Wanner, Geometric Numerical Integration, Springer Ser.
  Comput. Math. 31, Springer-Verlag, Berlin, 2006.

\bibitem{Reich99}
S.~Reich, Backward error analysis for numerical integrators, SIAM J. Numer.
  Anal. 36~(5) (1999) 1549--1570.

\bibitem{Skeel09}
R.~D. Skeel, What makes molecular dynamics work?, SIAM J. Sci. Comput. 31~(2)
  (2009) 1363--1378.

\bibitem{Bond07}
S.~D. Bond, B.~J. Leimkuhler, Molecular dynamics and the accuracy of
  numerically computed averages, Acta Numerica 16 (2007) 1--65.

\bibitem{BondThesis}
S.~D. Bond, Numerical methods for extended {H}amiltonian systems with
  applications in statistical mechanics, Ph.D. thesis, University of Kansas,
  Lawrence, KS, USA (2000).

\bibitem{Jorgensen83}
W.~L. Jorgensen, J.~Chandrasekhar, J.~D. Madura, R.~W. Impey, M.~L. Klein,
  Comparison of simple potential functions for simulating liquid water, J.
  Chem. Phys. 79~(2) (1983) 926--935.

\bibitem{Miller02}
T.~F. {Miller III}, M.~Eleftheriou, P.~Pattnaik, A.~Ndirango, D.~Newns, G.~J.
  Martyna, Symplectic quaternion scheme for biophysical molecular dynamics, J.
  Chem. Phys. 116~(20) (2002) 8649--8659.

\bibitem{Horn04}
H.~W. Horn, W.~C. Swope, J.~W. Pitera, J.~D. Madura, T.~J. Dick, G.~L. Hura,
  T.~Head-Gordon, Development of an improved four-site water model for
  biomolecular simulations: {TIP4P}-{E}w, J. Chem. Phys. 120~(20) (2004)
  9665--9678.

\bibitem{Engle05}
R.~D. Engle, R.~D. Skeel, M.~Drees, Monitoring energy drift with shadow
  {H}amiltonians, J. Comput. Phys. 206 (2005) 432–--452.

\bibitem{Wolf99}
D.~Wolf, P.~Keblinski, S.~R. Phillpot, J.~Eggebrecht, Exact method for the
  simulation of {C}oulombic systems by spherically truncated, pairwise $r^{-1}$
  summation, J. Chem. Phys. 110~(17) (1999) 8254--8282.

\bibitem{Zahn02}
D.~Zahn, B.~Schilling, S.~M. Kast, Enhancement of the {W}olf damped {C}oulomb
  potential: Static, dynamic, and dielectric properties of liquid water from
  molecular simulation, J. Phys. Chem. B 106~(41) (2002) 10725--10732.

\bibitem{Chialvo01}
A.~A. Chialvo, J.~M. Simonson, P.~T. Cummings, P.~G. Kusalik, On the
  determination of orientational configurational temperature from computer
  simulation, J. Chem. Phys. 114~(15) (2001) 6514--6517.

\bibitem{Hunenberger05}
P.~H. H{\"u}nenberger, Thermostat algorithms for molecular dynamics
  simulations, Adv. Polym. Sci. 173 (2005) 105--149.

\bibitem{Travis08}
K.~P. Travis, C.~Braga, Configurational temperature control for atomic and
  molecular systems, J. Chem. Phys. 128~(1) (2008) 014111.

\bibitem{Holian90}
B.~L. Holian, A.~J. {De Groot}, W.~G. Hoover, C.~G. Hoover, Time-reversible
  equilibrium and nonequilibrium isothermal-isobaric simulations with
  centered-difference {S}toermer algorithms, Phys. Rev. A 41~(8) (1990)
  4552--4553.

\bibitem{Bond99}
S.~D. Bond, B.~J. Leimkuhler, B.~B. Laird, The {N}os{\'e}-{P}oincar{\'e} method
  for constant temperature molecular dynamics, J. Comp. Phys. 151~(1) (1999)
  114--134.

\bibitem{FrenkelBook}
D.~Frenkel, B.~Smit, Understanding Molecular Simulation, 2nd Edition, Academic
  Press, New York, 2002.

\bibitem{Tuckerman92}
M.~Tuckerman, B.~J. Berne, G.~J. Martyna, Reversible multiple time scale
  molecular dynamics, J. Chem. Phys. 97~(3) (1992) 1990--2001.

\bibitem{Davidchack09}
R.~L. Davidchack, R.~Handel, M.~V. Tretyakov, Langevin thermostat for rigid
  body dynamics, J. Chem. Phys. 130~(23) (2009) 234101.

\bibitem{BondPrivate}
S.~D. Bond, private communication.

\bibitem{Joyce71}
D.~C. Joyce, Survey of extrapolation processes in numerical analysis, SIAM Rev.
  13~(4) (1971) 435--490.

\bibitem{AllenBook}
M.~P. Allen, D.~J. Tildesley, Computer Simulation of Liquids, Oxford University
  Press, Oxford, 1987.

\end{thebibliography}

\end{document}